\documentclass{edp-jp4}
\usepackage{epsfig}
\usepackage{graphicx}

\newcommand{\bea}{\begin{eqnarray}}
\newcommand{\eea}{\end{eqnarray}}
\newcommand{\be}{\begin{equation}}
\newcommand{\ee}{\end{equation}}
\newcommand{\rb}{\mathbf{r}}
\newcommand{\rzb}{\mathbf{r}_0}
\newcommand{\rpb}{\mathbf{r}'}
\newcommand{\kb}{\mathbf{k}}
\newcommand{\kpb}{\mathbf{k}'}

\begin{document}

\title{Simple theoretical tools for low dimension Bose gases}
\author{Yvan Castin}
\address{Laboratoire Kastler Brossel, ENS, 24 rue
Lhomond, 75 231 Paris, France}
\maketitle
\begin{abstract} 
We first consider an exactly solvable classical field model
to understand the coherence properties and the density
fluctuations of a one-dimensional (1D) weakly interacting
degenerate
Bose gas with repulsive interactions at temperatures
larger than the chemical potential.
In a second part, using a lattice model for the quantum field,
we explain how to carefully generalize the usual Bogoliubov
approach to study a degenerate and weakly interacting Bose gas 
in 1D, 2D or 3D in the regime of weak density fluctuations.
In the last part, using the mapping to an ideal Fermi gas in second
quantized formalism,
we calculate and discuss physically the density fluctuations and the
coherence properties of a gas of impenetrable bosons in 1D.
\end{abstract}

\section{A classical field model for the 1D weakly interacting Bose gas}
\subsection{Reminder: quantum theory for the ideal Bose gas}
\subsubsection{Second quantized formalism}
We consider the case of spinless, non-relativistic bosons of
mass $m$ moving in a space of spatial coordinates of dimension $d$
and
stored  either in a trapping potential $U(\rb)$ 
or in a cubic box of size $L$
with periodic boundary conditions. In this subsection the
bosons are not interacting. The grand canonical Hamiltonian
has then the following expression in second quantized formalism:
\be
\hat{H} = \int d\rb\, \hat{\psi}^\dagger(\rb)(h_0-\mu)\hat{\psi}(\rb).
\ee
$\mu$ is the chemical potential, and
the differential operator $h_0$ includes the kinetic energy operator and
the trapping potential $U(\rb)$:
\be
h_0= -\frac{\hbar^2}{2m}\Delta + U(\rb)
\ee
where $\Delta$ is the Laplacian operator in dimension $d$.
Note that $U\equiv 0$ in the spatially homogeneous case of a cubic box.
The field operator $\hat{\psi}(\rb)$ annihilates a boson
in point $\rb$ and obeys the usual bosonic commutation relations:
\be
[\hat{\psi}(\mathbf{r}), \hat{\psi}(\mathbf{r}')] = 0 ,
\ee
\be
[\hat{\psi}(\mathbf{r}), \hat{\psi}^{\dagger}(\mathbf{r}')] = \delta(\mathbf{r}-\mathbf{r}').
\ee

It is convenient to expand the field operator on the orthonormal basis of
the eigenmodes $\phi_\alpha(\rb)$ of $h_0$ with eigenenergy
$\epsilon_\alpha$:
\be
{\hat{\psi}}(\rb) = \sum_\alpha \phi_\alpha(\rb) \hat{a}_{\alpha}
\ee
\be
\left[-\frac{\hbar^2}{2m}\Delta + U(\rb) \right]\phi_\alpha(\rb) =
\epsilon_\alpha \phi_\alpha(\rb)
\ee
where $\hat{a}_{\alpha}$ annihilates a particle in mode $\phi_\alpha$.
The grand canonical Hamiltonian is then a sum of decoupled harmonic oscillators:
\be
\hat{H} = \sum_\alpha (\epsilon_\alpha-\mu) \hat{a}_{\alpha}^\dagger\hat{a}_{\alpha}.
\ee

\subsubsection{In thermal equilibrium}
We consider for convenience that the gas is in thermal equilibrium in the grand
canonical ensemble so that the density operator of the system reads
\be
\hat{\rho} = \Xi^{-1} e^{-\beta\hat{H}}
\ee 
where $\beta=(k_B T)^{-1}$ is inversely proportional to the temperature $T$.
As the canonical ensemble is more realistic, since no reservoir
of particles is present in usual experiments, one has to check for each particular
case at hand that the two ensembles are almost equivalent, a point
that we will address in \ref{subsubsec:qcohcorr}.

For the ideal Bose gas the density operator is a Gaussian in the field
components so that Wick's theorem, an elementary proof of which is given
in the appendix \ref{appen}, allows to express the expectation value
\be
\langle \hat{O}\rangle \equiv \mbox{Tr} [\hat{O}\hat{\rho}]
\ee
of an arbitrary observable $\hat{O}$
in terms of the mean occupation numbers of the eigenmodes:
\be
n_\alpha \equiv \langle \hat{a}_{\alpha}^\dagger\hat{a}_{\alpha}\rangle.
\ee
These occupation numbers are given by the Bose formula, also rederived
in the appendix \ref{appen}:
\be
n_\alpha = \frac{1}{e^{\beta(\epsilon_\alpha-\mu})-1} .
\ee
Assuming here for convenience that the ground state energy of $h_0$ vanishes,
which amounts to shifting the chemical potential, we see that the positivity
of $n_\alpha$ imposes the following range of variation of $\mu$:
\be
-\infty < \mu < 0.
\ee

In what follows one considers a temperature regime significantly different
from the zero temperature case, that is one assumes that there is a large
number of eigenmodes of energy less than $k_B T$. In the spatially
homogeneous case this implies that the size of the box is larger
than the thermal de Broglie wavelength:
\be
L\gg\lambda.
\label{eq:qtl}
\ee
One recalls the various regimes then encountered for $\mu$ increasing from
$-\infty$ to zero, in the present case of a non-degenerate ground mode:
\begin{itemize}
\item $|\mu|\gg k_B T$: {\bf non-degenerate regime}. All the occupation
numbers are much smaller than unity and are well approximated by the Boltzmann
formula:
\be
n_\alpha \simeq e^{-\beta(\epsilon_\alpha-\mu}) \ll 1.
\ee
Summing this expression over $\alpha$ relates the chemical potential
to the mean number $N$ of particles. In the case of the cubic box, and if one
replaces the sum by an integral, as allowed by condition (\ref{eq:qtl}),
this leads to
\be
e^{\beta\mu} \simeq \rho\lambda^d
\ee
where $\rho=N/L^d$ is the mean density and $\lambda$ is the thermal
de Broglie wavelength
\be
\lambda^2 = \frac{2\pi\hbar^2}{m k_B T}
\ee
so that we recover the condition for a non-degenerate regime
\be
\rho \lambda^d \ll 1.
\ee

\item {\bf degenerate regime:} $|\mu|\ll k_B T$ so that the ground mode
of the gas has a large occupation number:
\be
n_0 \simeq \frac{k_B T}{|\mu|} \gg 1
\ee
where we have expanded the exponential in the Bose law to first order
in its argument.
In this case several modes have a large occupation number, as soon as
$|\mu|\ll k_B T$, because condition (\ref{eq:qtl}) holds.

\item {\bf the ground mode is more populated than any other mode:}
this regime is reached when the ground mode is more populated that
the first excited mode:
\be
n_0 \simeq \frac{k_B T}{|\mu|} \gg n_1 \simeq \frac{k_B T}{|\epsilon_1-\mu|}
\ee
which results in
\be
|\mu|\ll\epsilon_1.
\label{eq:mu_vs_e1}
\ee
It is the right time to recall the phenomenon of saturation of the
total occupation of the excited modes, a key consequence of the Bose
law: for a given temperature the maximal value of the excited modes
population is
\be
N' \equiv \sum_{\alpha\neq 0} n_\alpha < N'_{\rm max} \equiv
\sum_{\alpha\neq 0} \frac{1}{e^{\beta\epsilon_\alpha}-1}
\ee
where one used the fact that each occupation number, being an increasing
function of the chemical potential, is bounded from above by its value
for $\mu=0$.
An important consequence of (\ref{eq:mu_vs_e1}) is that the saturation
of the excited modes population is quasi reached, as $\mu$ is negligible
compared to all $\epsilon_\alpha$ in the sum defining $N'_{\rm max}$:
\be
N' \simeq N'_{\rm max}.
\ee

\item {\bf Bose-Einstein condensation}: the ground mode has an occupation
number on the order of the total population of the excited levels.
One has then
\be
n_0 \sim N'_{\rm max}.
\ee
One recovers in particular the condition to reach Bose-Einstein
condensation applicable for a finite size system, that is
in the absence of thermodynamical limit \cite{Houches99},
\be
N > N'_{\rm max}
\ee
and the typical value of the chemical potential when a clear condensate is
formed, assuming $n_0 \sim N$ \cite{Diu}:
\be
|\mu| \sim \frac{k_B T}{N}.
\ee
\end{itemize}

\subsubsection{3D vs 1D in the thermodynamical limit}\label{subsubsec:versus}
We estimate the value of $N'_{\rm max}$ in a box of length much larger
than the thermal de Broglie wavelength $\lambda$:
\be
N'_{\rm max} = \sum_{\kb\neq\mathbf{0}} \frac{1}{\exp(\beta\hbar^2 k^2/2m)-1}
\ee
where the components of the wavevector $\kb$ are integer multiple of
$2\pi/L$.

In 3D, the condition $L\gg\lambda$ allows to approximate the sum by an integral:
\be
N'_{\rm max} \simeq \zeta(3/2) \left(\frac{L}{\lambda}\right)^3
\ee
where $\zeta$ is the Riemann Zeta function. One finds that the contribution 
to the integral is indeed dominated by wavevectors such that $\hbar^2k^2/2m \sim k_B T$
since the density of states cuts the low $k$ contribution and the Bose law cuts the high
$k$ contribution. So in 3D in the thermodynamic limit,
the gas is essential either in the non-degenerate 
regime or in the Bose condensed regime.

In 1D one cannot replace the sum over $k$ by an integral without getting an
infrared  divergent result for $N'_{\rm max}$: 
this is an indication that the low $k$ dominate.
The proper approximation to get the leading order contribution
in $L/\lambda$ is to replace the Bose law by its low $k$ behaviour:
\be
N'_{\rm max}\simeq \sum_{k\neq 0} \frac{k_B T}{\hbar^2k^2/2m}=
\frac{\pi}{3} \frac{L^2}{\lambda^2}
\ee
where we used $\sum_{n=1}^{+\infty} 1/n^2 = \pi^2/6$.
So in 1D in the thermodynamic limit, one can reach an arbitrarily high degeneracy
$\rho\lambda$ without forming a Bose-Einstein condensate.

\subsubsection{Correlation functions of the 1D Bose gas in the thermodynamic limit}
\label{subsubsec:qcohcorr}
The first order correlation function of the field operator $\hat{\psi}$ is defined
in 1D as
\be
g_1(z) \equiv \langle \hat{\psi}^{\dagger}(z) \hat{\psi}(0)\rangle.
\ee
It allows to determine the coherence length of the gas.
In a 1D box, $g_1$ is the Fourier transform of the momentum distribution:
\be
g_1(z) = \frac{1}{L}\sum_{k} n_{k} e^{i k z}
\rightarrow \int \frac{dk}{2\pi} n_k e^{i k z}
\ee
where $\rightarrow$ corresponds to the thermodynamic limit. In the non-degenerate regime
the momentum distribution reduces to the Boltzmann distribution so that $g_1$
is a Gaussian:
\be
g_1(z) \simeq \rho e^{-2\pi z^2/\lambda^2}.
\ee
In the degenerate regime, where $|\mu| \ll k_B T$, we replace the Bose law by its low
energy approximation, as we did in the calculation
of $N'_{\rm max}$, which results in a Lorentzian approximation to the momentum
distribution:
\be
n_k \simeq \frac{k_B T}{|\mu| +\hbar^2 k^2/2m}.
\ee
After integration over $k$ one obtains the equation of state of the gas:
\be
\rho\lambda = \left(\frac{\pi k_B T}{|\mu|}\right)^{1/2}
\ee
and after Fourier transform, one obtains an exponential function for $g_1$:
\be
g_1(z) \simeq \rho e^{-|z|/l_c}
\label{eq:g1_exp}
\ee
with the coherence length
\be
l_c = \frac{\rho \lambda^2}{2\pi} \gg \lambda.
\ee
Note that in 3D, when a Bose-Einstein condensate is present, $g_1$ tends
at infinity  to the condensate density $\rho_0$.

The second order correlation function of the field to be considered here is
\be
g_2(z) = \langle \hat{\psi}^\dagger(z)\hat{\psi}^\dagger(0)\hat{\psi}(0)
\hat{\psi}(z)\rangle.
\ee
Note that it is simply the spatial correlation function of the field intensity
(that is the density) for $z\neq 0$. It is also proportional to the probability
density, when measuring the positions of the particles, of detecting the first
particle in $0$ and the second one in $z$, as is most apparent in first
quantized form
\be
g_2(z)=  \sum_{i\neq j} \langle \delta(\hat{z}_j-z)\delta(\hat{z_i})\rangle
\ee
where $\hat{z}_i$ is the operator giving the
position of particle $i$ along $z$.

For an ideal Bose gas at thermal equilibrium in the grand canonical ensemble,
the second order correlation function is readily expressed in terms of $g_1$
through Wick's theorem:
\be
g_2(z) = \rho^2 + g_1^2(z)
\ee
where we used the fact that $g_1$ is real. This identity illustrates the phenomenon
of bosonic bunching in real space: $g_2$ is maximum in $z=0$ where it reaches 
twice its asymptotic value $\rho^2$. This implies strong density fluctuations
correlated over a spatial scale on the order of the coherent length $l_c$, 
a very negative fact from the atom laser perspective.

What happens in 3D? The same identity for $g_2$ holds, as a consequence of Wick's theorem.
When a condensate is present, one then finds also large density fluctuations
at an arbitrarily large scale: $g_2(+\infty) = \rho^2 +\rho_0^2 > \rho^2$. 
Fortunately, this prediction is not correct physically and is an artifact of the 
anomalously large fluctuations of the number of condensate particles in the grand
canonical ensemble for an ideal Bose gas. Wick's theorem implies
\be
\langle a_0^{\dagger 2} a_0^2\rangle = 2 n_0^2
\ee
where $a_0$ annihilates one particle in the condensate mode, so that
the standard deviation of the number of condensate particles is predicted
to be equal to its mean value $n_0$, which is not correct physically! 
As a consequence the fluctuations
of the total number of particles $N$ becomes also anomalously large, 
$\Delta N \sim \langle N\rangle$, and the equivalence between the grand canonical 
ensemble and the canonical ensemble is lost.
This problem is well known for the calculation of the fluctuations
of the Bose condensed ideal Bose gas 
\cite{Holthaus}. An easy way to avoid this problem is to
use the canonical ensemble, 
plus a number conserving Bogoliubov type approximation,
see section 7.8 of \cite{Houches99} or \cite{cartago}.

In 1D, when can we use the grand canonical ensemble to calculate
fluctuations?  In a box of size $L$, the variance of the total number of particles 
is related to the mean number of particles and to $g_2$ by
\be
\mbox{Var}N = \langle N\rangle + \int_0^L dz\, \int_0^L dz'\, [g_2(z-z')-\rho^2].
\ee
When no condensate is present in 1D, $L\gg l_c$ so that, using Wick's theorem
and the expression (\ref{eq:g1_exp}) for $g_1$, one gets
\be
\frac{\Delta N}{\langle N\rangle} \simeq \left(\frac{l_c}{L}
\right)^{1/2} \ll 1
\ee
which legitimates the use of the grand canonical ensemble.

\subsection{Construction of a classical field model}
The model of $N$ bosons in 1D interacting with a Dirac $\delta$ potential
is exactly solvable, see \cite{Lieb,Yang,Gaudin} 
for the case of repulsive interactions
and \cite{Herzog} for the attractive case.  However the calculation of the correlation
functions $g_1$ and $g_2$ is difficult, if one wants to 
access the full position dependence. A notable exception is the calculation
of $g_2(0)$, which is linked to the free energy thanks to the Hellman-Feynman
theorem and therefore requires only a knowledge of the energy spectrum
\cite{Drummond}.
We present here the classical field
version of this model, which is also exactly solvable and in a much easier way,
as done in  \cite{Scalapino,Atomlas}. We restrict here to the repulsive case.
The attractive case is indeed physically very different, as solitons can be
formed and the canonical ensemble, rather than the grand canonical one, has
to be used, which makes the classical field model more difficult to handle.

The intuitive idea to construct the classical field model is to replace
the quantum field operator $\hat{\psi}(z)$ by a complex field $\psi(z)$.
The Hamiltonian for the quantum field
\be
\hat{H}= \int dz\, \left[\frac{\hbar^2}{2m} (\partial_z\hat{\psi}^\dagger)
(\partial_z\hat{\psi}) + \frac{g}{2} \hat{\psi}^{\dagger2}
\hat{\psi}^2 -\mu \hat{\psi}^\dagger\hat{\psi}\right]
\ee
is then replaced by the energy functional of the classical field
\be
E[\{\psi\}] = \int dz\, \left[
\frac{\hbar^2}{2m} \left|\frac{d\psi}{dz}\right|^2 
+\frac{g}{2} |\psi|^4-\mu|\psi|^2
\right].
\label{eq:ener_func}
\ee
As a consequence the thermal density operator for the quantum field
is replaced by a probability distribution for the classical field:
\be
\rho=Z^{-1} e^{-\beta H} \rightarrow 
P_{\rm class}[\{\psi\}] = Z_{\rm class}^{-1} 
e^{-\beta E[\{\psi\}]}.
\ee
The first order correlation function $g_1$ is then given by the
average of $\psi^*(z)\psi(0)$ over $P_{\rm class}$, and $g_2$
is given by the expectation value of $|\psi|^2(z)|\psi|^2(0)$.

A more systematic way of constructing this classical field model
is to use the Glauber-P representation of the density operator,
a common tool of quantum optics \cite{Gardiner}, defined by the
identity:
\be
\hat{\rho} = \int {\cal D}\psi P[\{\psi\}]
|\{\psi\}\rangle
\langle\{\psi\}|
\ee
where the integral is a functional integral over all values of the complex
field $\psi$ and where we have introduced the Glauber coherent state of the matter field:
\be
|\{\psi\}\rangle\equiv e^{-\int |\psi|^2/2} e^{\int dz\, \psi(z)\hat{\psi}^\dagger(z)}
|0\rangle.
\ee
We recall that this coherent state, parameterized by the classical field $\psi(z)$,
is an eigenstate of $\hat{\psi}(z)$ with the eigenvalue $\psi(z)$, for all $z$.
We also note that the distribution $P$ is not necessarily positive, nor even
a regular function as it can be a distribution.
The expectation value of normally ordered expressions, that is with all the
$\hat{\psi}^\dagger$ on the left and all the $\hat{\psi}$'s on the right, are then exactly
related
to expectation values of classical field products with respect to the distribution
$P$. For example
\be
g_1(z) = \int {\cal D}\psi \, P[\{\psi\}]\, \psi^*(z)\psi(0).
\ee
The classical field model is therefore defined by the substitution
$P\rightarrow P_{\rm class}$.

The effect of this substitution is well monitored in the case of the ideal Bose
gas. The Glauber P distribution of the quantum field at thermal equilibrium is 
a Gaussian:
\be
P[\{\psi\}] \propto e^{-\sum_k |\alpha_k|^2/n_k}
\ee
where the sum is taken over the eigenmodes of the one-body Hamiltonian, that are
plane waves in the homogeneous case, $\alpha_k$ is the amplitude of the complex field $\psi$
on the mode $k$ and $n_k$ is the mean occupation number given by the Bose law \cite{normali}.
The field distribution $P_{\rm class}$ in the classical field model is also a Gaussian
and is obtained by the substitution
\be
n_k \rightarrow n_k^{\rm class} = \frac{k_B T}{\epsilon_k - \mu}
\ee
where $\epsilon_k$, here equal to $\hbar^2 k^2/2m$, is the eigenenergy of the
mode $k$. One then recovers the equipartition theorem 
\be
(\epsilon_k - \mu) n_k^{\rm class} = k_B T
\ee
as Boltzmann would have written it for a complex classical field
at thermal equilibrium.
We actually already used this expression for the occupation numbers, 
as a low energy approximation to the Bose law in the degenerate regime,
see \ref{subsubsec:versus}. The classical field formula is indeed close
to the Bose law for modes with a large occupation number: these modes of the field
have almost negligible quantum fluctuations \cite{Svistunov}.

When is the classical field model expected to give predictions close to the full
quantum theory? The answer depends on the observables and on the dimensionality of
space. One has to remember first that
the classical field model is subject to divergences in the absence of an energy cut-off,
the most famous illustration being the black body catastrophe of nineteenth century physics. 
One should first introduce an energy cut-off $\epsilon_{\rm max}$ on the energy of the modes, typically
on the order of $k_B T$ or larger.
Some observables depend on this cut-off, like the mean energy of the gas, so they cannot be calculated
with precision in the classical field model. Other observables, like the functions
$g_1$ and $g_2$ in 1D (and this is a nice feature of 1D), do not depend asymptotically
on the cut-off energy, as we have seen already for the ideal Bose gas and as is also the case
in presence of interactions. In the ideal Bose gas case, the validity condition of the classical
field approximation for the calculation of $g_1,g_2$ is, 
as we have seen,  $k_B T \gg |\mu|$, that is the gas has to be strongly degenerate.
The same condition holds in the weakly interacting case, though with a different equation of state
\cite{Atomlas}: a more detailed discussion is postponed to the
end of \S\ref{subsubsec:Bogclas}.
Finally we note that the shift in the 3D critical temperature of a Bose gas due to interactions
was calculated recently with a very high precision using a classical field model
\cite{Svistunov2,Moore}.

\subsection{Solution of the classical field model}

\subsubsection{The problem to be solved}
One has to be able to calculate functional integrals of the type
\be
g_1(z) = \frac{\int {\cal D}\psi \, \psi^*(z)\psi(0) e^{-\beta E[\psi]}}
{\int {\cal D}\psi \, e^{-\beta E[\psi]}}
\ee
where the complex field $\psi$ satisfies the periodic boundary conditions
$\psi(0)=\psi(L)$ \cite{intfunc}.
The technique to implement this periodicity is to integrate
separately over all the possible values $\psi_0$ of $\psi$ in $z=0$ and over all
possible closed paths starting in $\psi_0$ (for $z=0$) and ending in
$\psi_0$ (for $z=L$). In the same spirit, in the numerator giving $g_1$, one treats
also separately the intermediate point of coordinate $z$, by integrating over
the complex value $\psi_z$ of the field in $z$. As a consequence
\bea
 \int {\cal D}\psi \, \psi^*(z)\psi(0) e^{-\beta E[\psi]} = \nonumber\\
 \int d\psi_0 \int d\psi_z 
\left[\int_{\psi(0)=\psi_0}^{\psi(z)=\psi_z} {\cal D}\psi_{0\rightarrow z}\,
e^{-\beta E_{0\rightarrow z}} \right]
\left[\int_{\psi(z)=\psi_0}^{\psi(L)=\psi_0} {\cal D}\psi_{z\rightarrow L}\,
e^{-\beta E_{z\rightarrow L}} \right]
\psi^*(z)\psi(0) 
\label{eq:numerator}
\eea
where $\int d\psi_0$ stands for the integral over the real part and the imaginary part
of $\psi_0$ from $-\infty$ to $+\infty$.
One is then left with the calculation of functional integrals of the type
\be
\int_{\psi(z_i)=\psi_i}^{\psi(z_f)=\psi_f}
{\cal D}\psi_{z_i\rightarrow z_f}\, e^{-\beta E_{z_i\rightarrow z_f} }
\ee
that is over possible complex fields defined over the spatial interval $(z_i,z_f)$
with fixed values $\psi_{i,f}$ at the end points of the interval. The energy
of a field configuration over this interval is the spatial integral of the energy density
\be
E_{z_i\rightarrow z_f} = \int_{z_i}^{z_f} dz\, 
\left[
\frac{\hbar^2}{2m} \left|\frac{d\psi}{dz}\right|^2 
+\frac{g}{2} |\psi|^4-\mu|\psi|^2
\right].
\ee

\subsubsection{Reminder: Feynman's formulation of quantum mechanics}
Consider a quantum particle of mass $M$ moving in 2D in a static
potential $V(x,y)$. The corresponding Hamiltonian is
\be
\hat{\cal H} = \frac{\hat{p}_x^2+\hat{p}_y^2}{2M}
+ V(\hat{x},\hat{y})
\ee
where $\hat{p}_x,\hat{p}_y,\hat{x},\hat{y}$ are the momentum and the position
operators along each direction of space.
The so-called Feynman propagator in real time can be written in terms of a functional
integral over two real valued paths $x(t),y(t)$:
\be
\langle x_f,y_f | e^{-i{\cal H}(t_f-t_i)/\hbar}|x_i,y_i\rangle
= \int_{x(t_i)=x_i}^{x(t_f)=x_f} {\cal D}x
  \int_{y(t_i)=y_i}^{y(t_f)=y_f} {\cal D}y
e^{i\int_{t_i}^{t_f} dt\, {\cal L}/\hbar}
\ee
where the Lagrangian is
\be
{\cal L} = \frac{1}{2} M \left[
\left(\frac{dx}{dt}\right)^2+\left(\frac{dy}{dt}\right)^2
\right]-V[x(t),y(t)].
\ee

We shall need this formulation of quantum mechanics for the propagator in
imaginary time. Inside the integrals one performs the substitutions
\bea
i t_f &\rightarrow& \tau_f \\
i t_i &\rightarrow& \tau_i \\
i t &\rightarrow& \tau \\
i dt &\rightarrow& d\tau 
\eea
where all the `times' $\tau$'s are real, to obtain 
\be 
\langle x_f,y_f | e^{-{\cal H}(\tau_f-\tau_i)/\hbar}|x_i,y_i\rangle
= \int_{x(\tau_i)=x_i}^{x(\tau_f)=x_f} {\cal D}x
  \int_{y(\tau_i)=y_i}^{y(\tau_f)=y_f} {\cal D}y
e^{-\int_{\tau_i}^{\tau_f} d\tau\, {\cal E}/\hbar}
\ee
which involves now the energy rather than the Lagrangian:
\be
{\cal E} = \frac{1}{2} M \left[
\left(\frac{dx}{d\tau}\right)^2+\left(\frac{dy}{d\tau}\right)^2
\right]+V[x(\tau),y(\tau)].
\ee

\subsubsection{Quantum reformulation}
We map the 1D classical field model at thermal equilibrium to a fictitious quantum
propagator in imaginary time of a particle moving in a two-dimensional space,
with the help of the following correspondence table.

\bigskip
\begin{center}
\begin{tabular}{|c|c|}
\hline 
classical field model & quantum analogy \\
\hline
path integral & quantum propagator in imaginary time \\
abscissa $z$ & `time' $\tau$ \\
Re\,$\psi(z)$ & $x(\tau)$ \\
Im\,$\psi(z)$ & $y(\tau)$ \\
$\int_{\psi(z_i)=\psi_i}^{\psi(z_f)=\psi_f}
{\cal D}\psi_{z_i\rightarrow z_f}\, e^{-\beta E_{z_i\rightarrow z_f}} $
& 
$\langle x_z,y_z | e^{-{\cal H}(\tau_f-\tau_i)/\hbar}|x_i,y_i\rangle$ \\
\hline
\end{tabular}
\end{center}
\bigskip

This correspondence is exact provided that $\beta$ times the energy density of the classical field
transforms into the energy ${\cal E}$ over $\hbar$, which imposes the mass $M$ of the
fictitious particle
\be
M = \frac{\hbar^3}{m k_B T}
\ee
and the fictitious external potential of the 2D quantum problem
\be
V(x,y) = \hbar\beta\left[\frac{g}{2}(x^2+y^2)^2-\mu(x^2+y^2) \right].
\ee

The numerator of $g_1$, given in (\ref{eq:numerator}),
is transformed as follows with the quantum mechanical analogy:
\bea
&& \int\! d\psi_0 \int\! d\psi_z \left[\int_{\psi(0)=\psi_0}^{\psi(z)=\psi_z} {\cal D}\psi_{0\rightarrow z}\,
e^{-\beta E_{0\rightarrow z}} \right]
\left[\int_{\psi(z)=\psi_0}^{\psi(L)=\psi_0} {\cal D}\psi_{z\rightarrow L}\,
e^{-\beta E_{z\rightarrow L}} \right]
\psi^*(z)\psi(0) \nonumber \\
&=& \int\! dx_0\!\int\! dy_0\! \int\! dx_z\!\int\! dy_z  \langle x_z,y_z|e^{-z{\cal H}/\hbar}
|x_0,y_0\rangle\langle x_0,y_0 |e^{-(L-z){\cal H}/\hbar} |x_z,y_z\rangle
(x_z-iy_z) (x_0+iy_0) \nonumber \\
&=& \mbox{Tr}\left[e^{-(L-z){\cal H}/\hbar}(\hat{x}-i\hat{y}) e^{-z{\cal H}/\hbar}(\hat{x}+i\hat{y})
\right]. \ \ \ \ \ \ \ 
\eea
The denominator of $g_1$ is then simply the trace of $\exp(-L\hat{\cal H}/\hbar)$.

The calculation of the trace is performed in practice in the eigenbasis of ${\cal H}$.
As this Hamiltonian is rotationally invariant, one classifies its eigenstates with 
the angular momentum quantum number $l$, integer from $-\infty$ to $+\infty$, and 
the radial quantum number $n$, integer from $0$ to $+\infty$. The corresponding normalized 
eigenvector is denoted $|\phi_n^l\rangle$.
For a given $l$, the eigenenergies are sorted in increasing order starting from $n=0$.
The absolute ground state is obtained for $l=0$ since a non-vanishing angular momentum
gives rise to a centrifugal energy term that can only increase the energy.
A great simplification occurs in the thermodynamical limit, that is when
\be
L(\epsilon_{\rm exc}-\epsilon_0^{l=0})/\hbar \gg 1
\ee
where $\epsilon_{\rm exc}$ is the lowest energy eigenvalue above the ground
state energy $\epsilon_0^{l=0}$. In this case one can approximate the imaginary
time evolution operator by restricting to the lowest energy mode(s), e.g.
\be
e^{-L{\cal H}/\hbar} \simeq
e^{-L\epsilon_0^{l=0}/\hbar} |\phi_0^{l=0}\rangle\langle\phi_0^{l=0}|.
\ee
In the thermodynamic limit, one therefore obtains the expressions
\bea
\label{eq:g1c}
g_1(z) &=& \langle\phi_0^{l=0}|(\hat{x}-i\hat{y}) e^{-z{(\cal H}-\epsilon_0^{l=0})/\hbar}
(\hat{x}+i\hat{y})|\phi_0^{l=0}\rangle \\
g_2(z) &=& \langle\phi_0^{l=0}|(\hat{x}^2+\hat{y}^2) e^{-z{(\cal H}-\epsilon_0^{l=0})/\hbar}
(\hat{x}^2+\hat{y}^2)|\phi_0^{l=0}\rangle.
\label{eq:g2c}
\eea

\subsubsection{Results}\label{subsubsec:results}
We briefly present the predictions of the classical field model, more details being
given in \cite{Atomlas}.

A first point is to realize that, with the  proper system of units, the state of the classical
field in the thermodynamic limit is controlled by a single dimensionless parameter. If one expresses $\psi$
in units of the square root of the mean density $\rho$ and the length $z$ in units of the coherence
length of the ideal Bose gas,
\be
z = \frac{\hbar^2\rho}{m k_B T} Z = \frac{\rho\lambda^2}{2\pi} Z
\ee
where $Z$ is dimensionless, and if one realizes that the chemical potential
$\mu$ is fixed by the condition 
\be
\langle |\psi|^2\rangle = \rho
\ee
one can check that the classical field energy functional
$E$ in (\ref{eq:ener_func}) depends on the single parameter
\be
\chi = \frac{\hbar^2\rho^2}{mk_B T} \frac{\rho g}{k_B T}.
\label{eq:def_chi}
\ee

\begin{figure}[htb]
\centerline{\includegraphics[height=6cm,clip]{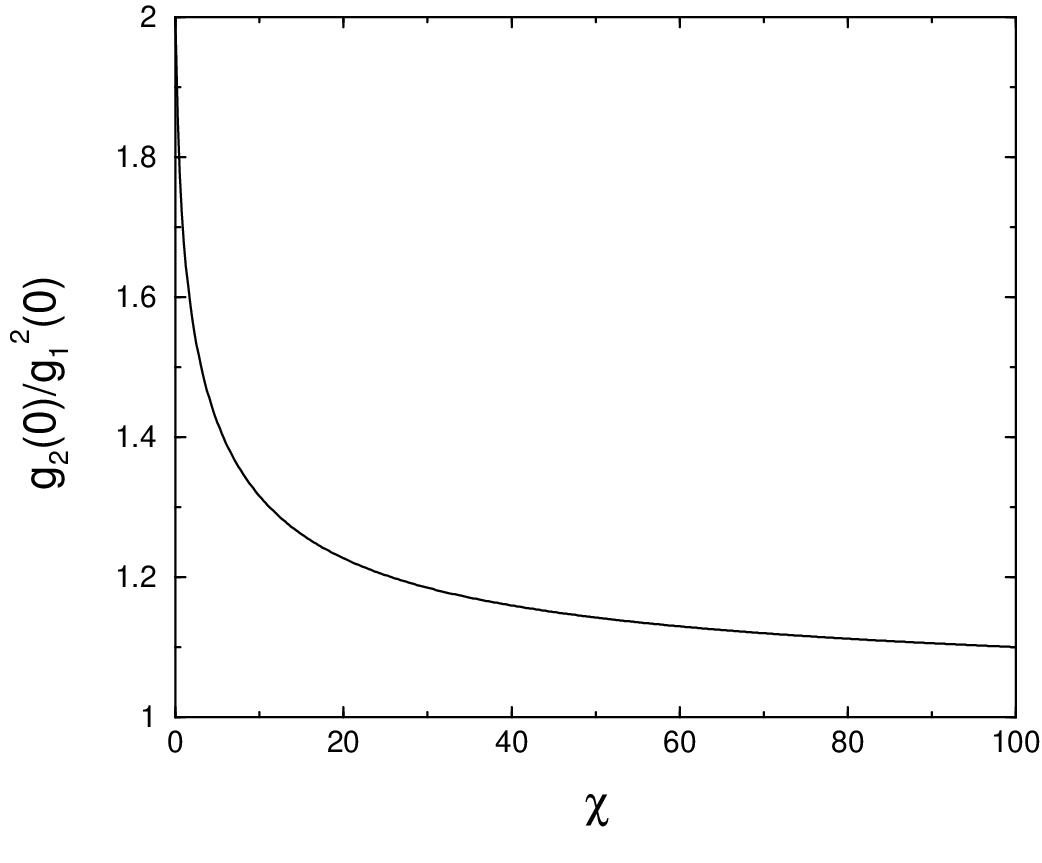}}
\caption{
In the classical field model, contrast $C$ of the local density fluctuations of a 
1D Bose gas as function of the temperature dependent interaction strength $\chi$ 
defined in Eq.(\ref{eq:def_chi}).
\label{fig:contrast}}
\end{figure}

Our concern for atom laser applications was the large intensity fluctuation of the
ideal Bose gas. In the classical field model in the thermodynamic limit, 
the function $g_2$ is a decreasing function
of $z \in (0,+\infty)$, since ${\cal H}-\epsilon_0^{l=0}$ is positive. It reaches its maximum
value in $z=0$ so we introduce the contrast
\be
C\equiv \frac{g_2(0)}{g_1^2(0)}= 
\frac
{\langle\phi_0^{l=0}|(\hat{x}^2+\hat{y}^2)^2|\phi_0^{l=0}\rangle}
{\langle\phi_0^{l=0}|\hat{x}^2+\hat{y}^2|\phi_0^{l=0}\rangle^2}.
\ee
This contrast is plotted as function of $\chi$ in figure \ref{fig:contrast}.
For $\chi=0$, one recovers the ideal Bose gas value $C=2$. When $\chi$ increases,
there is a very sharp decrease of $\chi$, followed by a power low tail, $C$ tending
slowly to unity when $\chi$ tends to $+\infty$. The curve shows that repulsive interactions
efficiently reduce the density fluctuations of the gas.

\begin{figure}[htb]
\centerline{\includegraphics[height=6cm,clip]{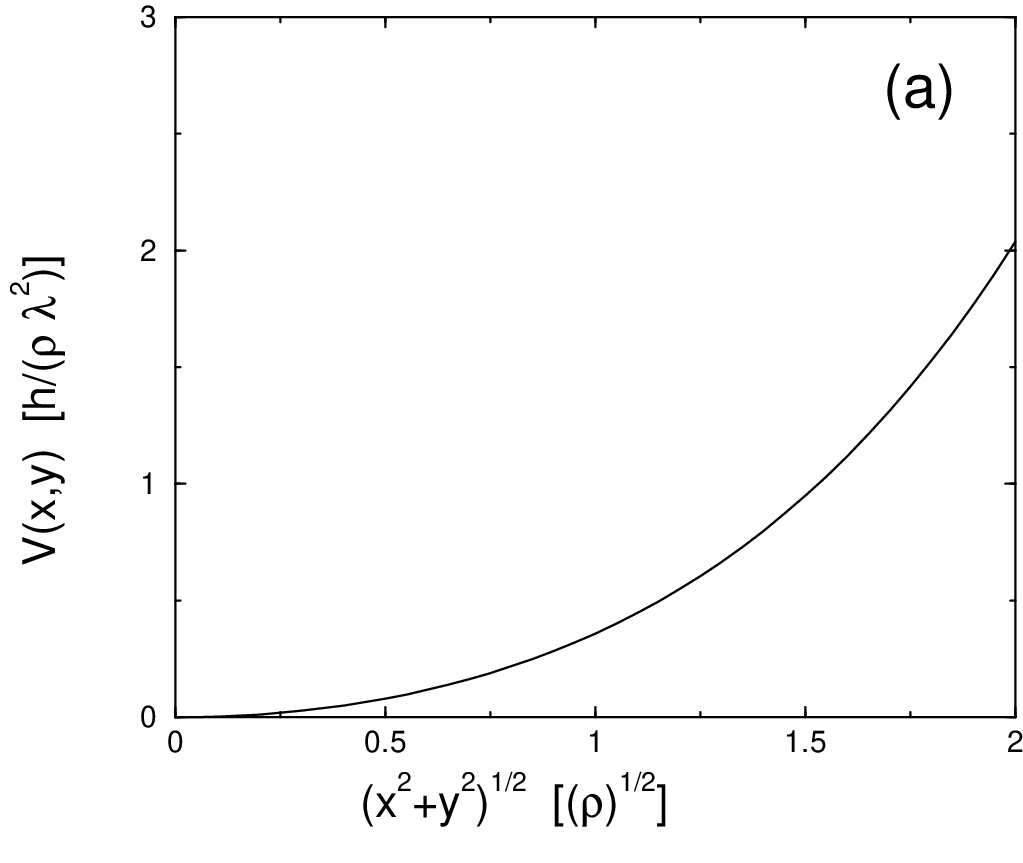}
\hspace{1cm}
\includegraphics[height=6cm,clip]{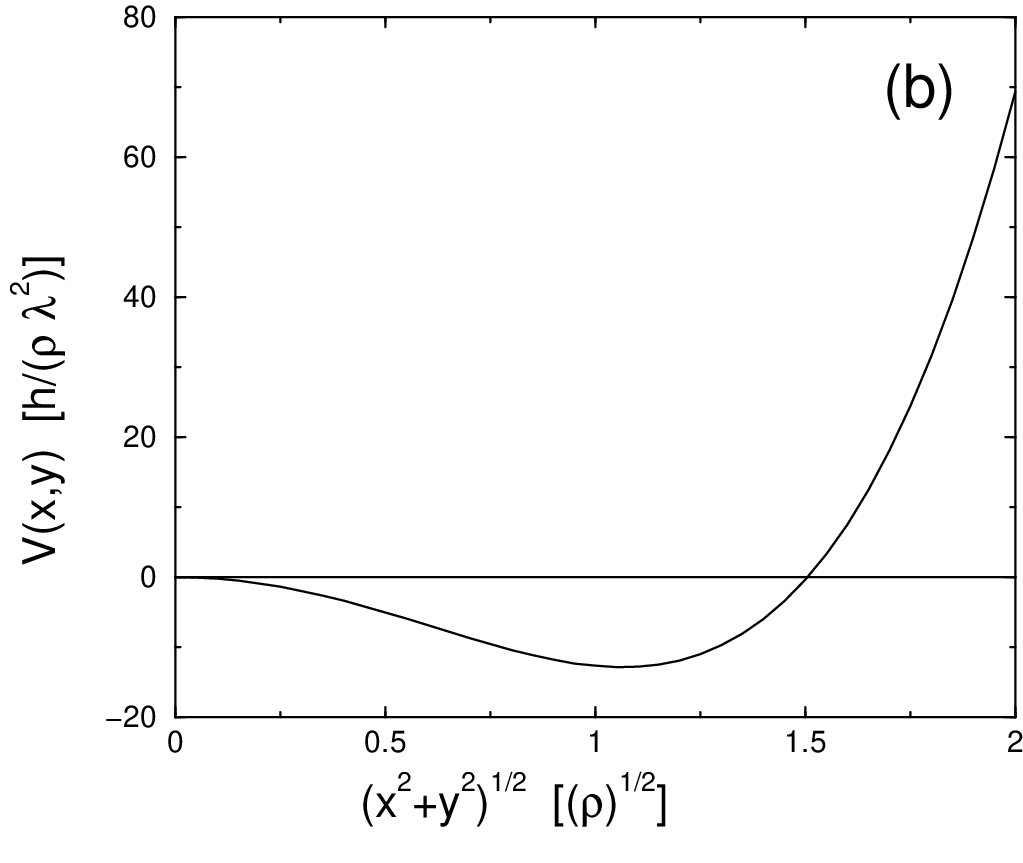}}
\caption{
In the quantum mechanics equivalent to the classical field problem, potential $V(x,y)$ seen by the
quantum mechanical particle.
(a) For a low value of $\chi=0.1$ and (b) for a large value of $\chi=20$.
The units on the axes are such that $V$ depends only on the parameter $\chi$ 
defined in Eq.(\ref{eq:def_chi}).
\label{fig:pot}}
\end{figure}

What is the physical origin of the very fast decrease of $C$ in the vicinity of
$\chi=0$? This sharp crossover from strong to weak density fluctuations corresponds
to a change of shape of the fictitious potential $V$. When $\chi<\chi_c\simeq 0.28$,
the chemical potential is negative so that $V$ is minimum in $x=y=0$, see figure \ref{fig:pot}a,
and the field remains trapped around $\phi=0$, with large intensity fluctuations.
When $\chi>\chi_c$,
$\mu$ is positive so that $V$ is a Mexican hat potential, see figure \ref{fig:pot}b,
and the field is trapped around a non-vanishing value of its modulus, with weak density
fluctuations.

The large $z$ behaviour of the correlation functions is also easy to access.
In (\ref{eq:g1c}), $\hat{x}+i\hat{y}$ maps $|\phi_0^{l=0}\rangle$ onto a $l=1$
state so it is sufficient to inject a closure relation on the eigenstates of
${\cal H}$ within the $l=1$ manifold. When $z\rightarrow+\infty$
one can keep the contribution of the mode $n=0,l=1$ so that $g_1$ is proved
to vanish exponentially with $z$:
\be
g_1(z)  \simeq a_1 e^{-\kappa_1 z} \ \ \ \mbox{with} \ \ \ 
\kappa_1 = \frac{\epsilon_0^{l=1}-\epsilon_0^{l=0}}{\hbar}
\ee
where $a_1 = |\langle\phi_0^{l=1}|\hat{x}+i\hat{y}|\phi_0^{l=0}\rangle|^2$
does not depend on $z$. We may therefore define the coherence length of the gas
as $1/\kappa_1$.
For $g_2$ the closure relation is in the $l=0$ manifold and one keeps the contribution
of the ground and the first excited mode in the large $z$ limit:
\be
g_2(z) \simeq g_1^2(0) + a_2 e^{-\kappa_2 z} \ \ \ \mbox{with} \ \ \ 
\kappa_2 = \frac{\epsilon_1^{l=0}-\epsilon_0^{l=0}}{\hbar}
\ee
where $a_2 = |\langle\phi_1^{l=0}|\hat{x}^2+\hat{y}^2|\phi_0^{l=0}\rangle|^2$
is a constant. The correlation length of the gas may be defined as $1/\kappa_2$.

\begin{figure}[htb]
\centerline{\includegraphics[height=6cm,clip]{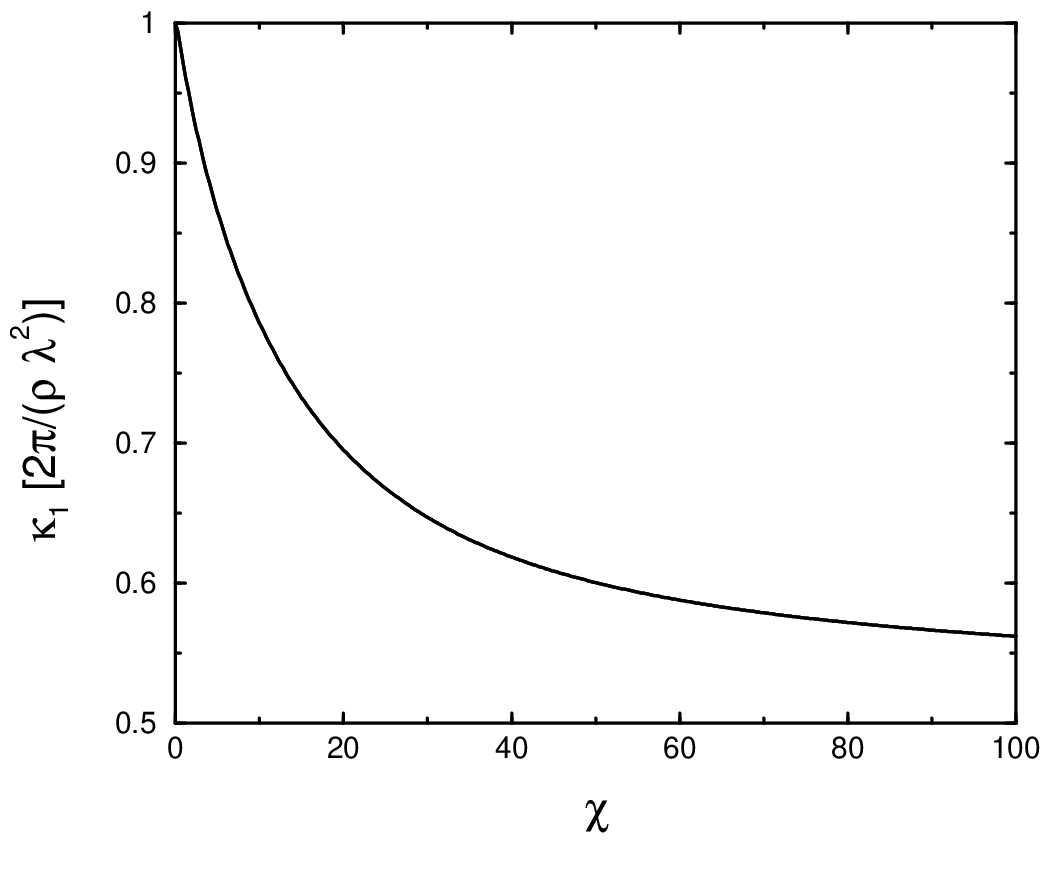}
\hspace{5mm}
\includegraphics[height=6cm,clip]{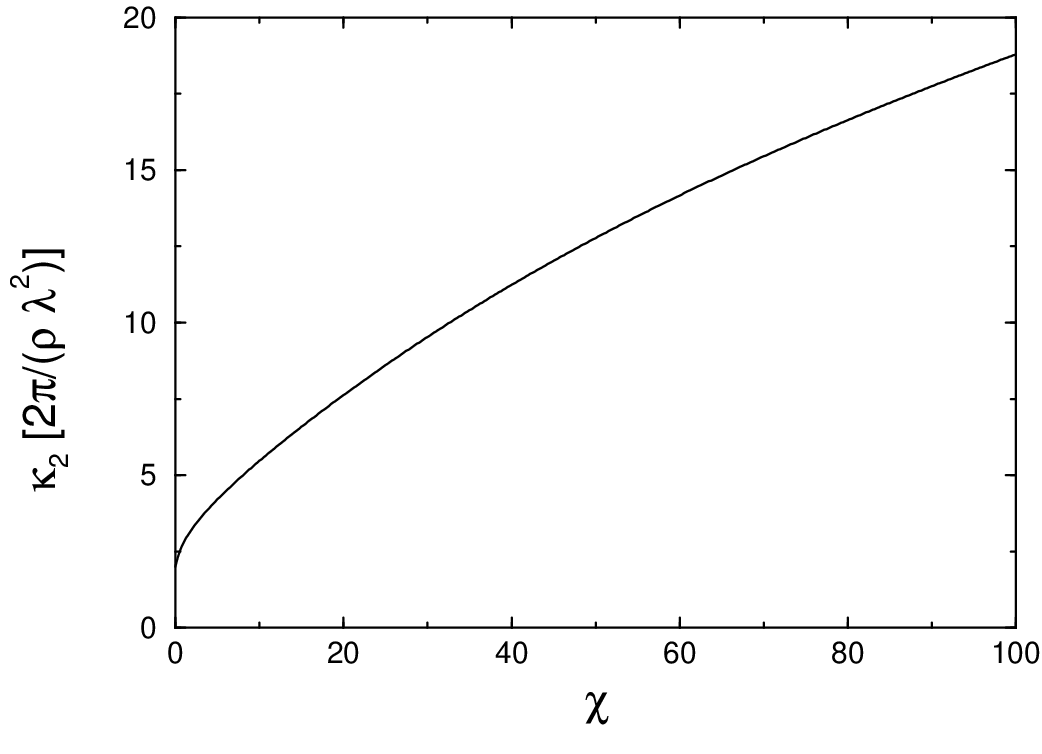}}
\caption{In the one-dimensional classical field model, the inverse field coherence length $\kappa_1$ and
the inverse intensity correlation length $\kappa_2$ as function of the temperature dependent
interaction parameter $\chi$ defined in Eq.(\ref{eq:def_chi}).
\label{fig:cohcorr}}
\end{figure}

The values of $\kappa_1, \kappa_2$ are plotted in figure \ref{fig:cohcorr}.
For $g=0$ one recovers the ideal Bose gas results of \ref{subsubsec:qcohcorr}.
For repulsive interactions, the coherence length is slightly increased, by a factor
of up to 2. The correlation length strongly decreases and its ratio with the coherence
length tends to zero in the large $\chi$ limit: the density fluctuations not only decrease
but also take place on a length scale much smaller than the coherence length of
the atom laser.

Analytical results can be obtained for $\kappa_1$ and $\kappa_2$ in the large $\chi$
limit by using the quantum mechanical analogy. Writing the wavefunction with angular
momentum $l$ as
\be
\phi^{l}(x,y) = \frac{f(r)}{\sqrt{2\pi r}}\, e^{i l \theta}
\ee
where $r$ and $\theta$ are polar coordinates in the $x,y$ plane, one has to solve
Schr\"odinger's equation
\be
-\frac{\hbar^2}{2M}f''(r) + U(r) f(r) = \epsilon f(r)
\ee
with the effective potential
\be
U(r) = V(r) +\frac{\hbar^2}{2Mr^2}(l^2-1/4)
\ee
and the boundary condition $f(0)=0$. Note the presence of a centripetal potential
for $l=0$, which does not support any bound state, a known peculiarity of the 2D quantum motion. 
In the large $\chi$ limit one finds the minimum of $U(r)$ in $r_0>0$ and one approximates
$U(r)$ around $r_0$ by a series expansion in powers of $r-r_0$. To leading order, $U(r)$ is
then approximated by a quadratic potential centered in $r_0$, 
then cubic corrections can be taken into account perturbatively, and so on.
One has to be careful in this type of expansion to collect all the terms of a given order.
For example, the calculation of the mean density proceeds through the identity
\be
\langle r^2\rangle = r_0^2 + 2\langle r-r_0\rangle r_0 +
\langle(r-r_0)^2\rangle
\ee
where the expectation value is taken in $|\phi_0^{l=0}\rangle$. One has then to take care
of the fact that the last two terms are of the same leading order, the last term being non-zero
already for the harmonic approximation to $U(r)$ 
whereas the other term $\langle r-r_0\rangle$
is first non-zero when the cubic distortion of $U(r)$ is included.
One obtains in the large $\chi$ limit:
\bea
\label{eq:eos}
\mu &=& \rho g \left[1+\frac{1}{2\chi^{1/2}}+\ldots\right] \\
\label{eq:k1_large_chi}
\kappa_1^{-1} &\simeq& \frac{\rho\lambda^2}{\pi}=2 l_c ^{\rm ideal} \\
\kappa_2^{-1} &\simeq& \frac{1}{2}\left(\frac{\hbar^2}{m\mu}
\right)^{1/2}= \frac{\xi}{2} \\
C &=&  1+\frac{1}{\chi^{1/2}}+\ldots\label{eq:C_large_chi}
\eea
where $\xi$ is the so-called healing length. The equation of state  (\ref{eq:eos})
exactly coincides with the Gross-Pitaevskii equation in the large $\chi$ limit,
and $\xi$ also naturally arises from the Gross-Pitaevskii equation. This may be 
surprising at first sight, since the Gross-Pitaevskii equation holds for the
condensate wavefunction, whereas there is no true condensate in 1D in the 
thermodynamic limit. This fact will be recover and hopefully will become clear
in section \ref{sec:Bogo} of this lecture.
We note that $\chi$ is proportional to $(l_c/\xi)^2$, 
so that the regime of weak density fluctuations corresponds to
$l_c \gg \xi$.

\subsubsection{A Bogoliubov approach for $\chi\gg 1$ \label{subsubsec:Bogclas}}
A limitation of the previous analytical expansion in the large $\chi$ limit
is that it is difficult to transpose to the quantum problem. Is there another approach
available that would apply both to the classical field and the quantum field?

One could think of applying the Bogoliubov approach to the classical field model.
A problem appears in the large $L$ limit since no condensate is present, as we now
show. In the classical field version of the Bogoliubov approach, one splits the
field as
\be
\psi(z) = N_0^{1/2} e^{i\theta}\frac{1}{L^{1/2}}+\psi_{\perp}(z)
\ee
where $\psi_{\perp}$ is the component of the field on the plane waves with
non-zero momentum and is supposed to be much smaller than the component 
on the $k=0$ wave, $\propto \sqrt{N_0/L}$, so that one can neglect all the
terms of degree $>2$ in 
$\psi_\perp$ in the energy function (\ref{eq:ener_func}).
The resulting quadratic function of the field can then be diagonalized
by the Bogoliubov transformation.
As shown in the lecture by Gora Shlyapnikov in this school,
one then has
\be
\int_0^L dz\, \langle |\psi_\perp(z)|^2\rangle = \sum_{k\neq 0}
(U_k^2+V_k^2) \frac{k_B T}{\epsilon_k^{\rm bog}}
\ee
where $U_k,V_k$ are the dimensionless amplitudes of the Bogoliubov modes on the
plane waves, normalized such that $U_k^2-V_k^2=1$, 
and $\epsilon_k^{\rm bog}$ is the usual Bogoliubov spectrum.
This sum is dominated by the low momenta:
\be
(U_k^2+V_k^2) \frac{1}{\epsilon_k^{\rm bog}} \sim_{k\rightarrow 0}
\frac{m}{\hbar^2 k^2}.
\ee
As a consequence the relative mean weight of the excited plane waves is
\be
\frac{\int |\psi_\perp|^2}{\int |\psi|^2} \simeq
\frac{mk_B T}{N\hbar^2} \sum_{k\neq 0} \frac{1}{k^2}= \frac{\kappa_1 L}{6}
\ee
where the inverse coherence length is given by (\ref{eq:k1_large_chi}).
We see that the basic assumption of Bogoliubov theory fails when $L\gg \kappa_1^{-1}$,
that is in the absence of a condensate.

The appropriate small parameter in the large $\chi$ domain is not the non-condensed
fraction but the relative density fluctuations. This gives the idea of using 
the intensity-phase representation of the field, an idea extensively developed
in \cite{Popov}. One writes
\be
\psi(z) = \rho^{1/2}(z) e^{i\theta(z)},
\ee
one recalculates the energy functional of (\ref{eq:ener_func}):
\be
E\left[\{\psi\}\right] =
\int_0^L dz \frac{\hbar^2}{2m}\left[
\left(\frac{d}{dz}\sqrt{\rho}\right)^2+
\rho \left(\frac{d\theta}{dz}\right)^2
+\frac{g}{2}\rho^2-\mu\rho \right].
\ee
One then introduces the density deviation $\delta\rho$ such that
\be
\rho(z) = \rho_0 + \delta\rho(z)
\ee
and $\rho_0$ is the constant such that the uniform field $\psi_0=\rho_0^{1/2}$
has the absolute minimum of energy:
\be
\mu = g\rho_0.
\ee
Finally one  quadratizes the energy functional in terms of $\delta\rho$ and $\theta$.
Remarkably this does not require weak phase fluctuations, only low density
fluctuations, so that
\be
\rho \left(\frac{d\theta}{dz}\right)^2 \simeq
\rho_0 \left(\frac{d\theta}{dz}\right)^2,
\ee
and this is a crucial point in the absence of a condensate!
This leads to the quadratic energy functional
\be
E[\{\psi\}]=\mbox{ct}+\int_0^L dz\, \left[\frac{\hbar^2}{8m}
\left(\frac{d\delta\rho}{dz}\right)^2+\frac{\hbar^2}{2m}\rho_0
\left(\frac{d\theta}{dz}\right)^2
+\frac{g}{2}(\delta\rho)^2
\right].
\ee
This energy functional may be put in normal form
by a Bogoliubov transformation.  If one imposes periodic boundary conditions
for the phase $\theta$, one obtains exactly the usual Bogoliubov
spectrum. This allows to recover the large $\chi$ analytical results
of \ref{subsubsec:results} \cite{winding_number}.

This modified Bogoliubov approach can tell when the classical field model
gives predictions close to the full quantum theory for $\chi\gg 1$:
\begin{itemize}
\item the shortest length scale obtained from the $g_1, g_2$ functions is the 
correlation length $\kappa_2^{-1}\sim \xi$, corresponding to a Bogoliubov
energy of the order of $\mu$. The corresponding Bogoliubov modes should have
a large occupation number for the quantum field to be approximated by a
classical field, which imposes 
\be
\label{eq:cond}
k_B T \gg \mu\simeq \rho g.
\ee
\item the gas should be degenerate. Since $(\rho\lambda)^2/2\pi = \chi k_B T/\rho g$,
the condition $\rho\lambda \gg 1$ is automatically satisfied for a large $\chi$
when (\ref{eq:cond}) holds.
\item Also a classical field model is realistic in the weakly interacting regime only:
strong correlations between the particles are absent in a classical field model 
$\ldots$ since there is no such a thing like a particle! This is apparent from the 
equation of state $\mu=\rho g$, which makes sense quantum mechanically in the
weakly interacting regime only, that is when $\rho \xi \gg 1$.
Since $\rho \xi = \chi^{1/2} k_B T/\rho g$, $\rho \xi$ is automatically
big in the large $\chi$ regime when (\ref{eq:cond}) holds.
\end{itemize}

\section{Extension of Bogoliubov method to quasi-condensates in the weakly interacting regime}
\label{sec:Bogo}

The 1D classical field model of the previous section has identified an interesting
regime for guided atom optics and atom lasers, a degenerate regime where the density fluctuations
are strongly reduced by the repulsive interactions between the particles and where the coherence
length is much larger than the thermal de Broglie wavelength. 
Such a state of the gas exists even if there is no Bose-Einstein condensate; it is called
a `quasi-condensate', and was extensively studied in \cite{Popov} for the
uniform case (see the lecture of Gora Shlyapnikov for the trapped case).

The goal of this section is to present a very simple but accurate theoretical frame, simpler than the
one of \cite{Popov}, to study the quasicondensates in the degenerate and weakly interacting regime.
Degenerate means that
\be
\rho \lambda^d \gg 1
\label{eq:degen}
\ee
where $\rho$ is the gas density, $\lambda$ is the thermal de Broglie wavelength and $d$
is the dimension of space. Weakly interacting means
\be
\rho \xi^d \gg 1
\label{eq:weak}
\ee
where the healing length $\xi$ is related to the chemical potential of the gas
by
\be
\mu = \frac{\hbar^2}{m\xi^2}.
\label{eq:def_xi}
\ee
Our theoretical frame is simply the Bogoliubov approach but in the density-phase
representation of the quantum field. In case where a condensate is present,
it gives results equivalent to the usual Bogoliubov approach.

\subsection{Construction of an appropriate model}
A difficulty often appearing in theoretical treatments of
the quasi-condensate problem is the appearance of divergences,
an infrared divergency in 1D, an infrared and an ultraviolet divergency in 2D.
We construct here with some care a model Hamiltonian that, when combined
with a systematic expansion in powers of small parameters, will allow
to avoid all these divergencies.

\subsubsection{General idea of the theory of quasi-condensates}
This general idea is for example inspired by the Bogoliubov approach
developed in the previous section for the classical field model.
One identifies as the small parameter the relative density fluctuations
of the gas and one performs a systematic expansion in this small parameter.
So one first writes the field operator as
\be
\hat{\psi}(\rb) = e^{i\hat{\theta}(\rb)}\sqrt{\hat{\rho}(\rb)}
\ee
where the operator giving the density is
\be
\hat{\rho}(\rb) = \hat{\psi}^\dagger(\rb)\hat{\psi}(\rb)
\ee
and where $\hat{\theta}$ is the mythical phase operator in point $\rb$.
Theses two operators are expected to be canonically conjugated:
\be
\label{eq:conj}
[\hat{\rho}(\rb),\hat{\theta}(\rpb)]= i \delta(\rb-\rpb).
\ee
One then splits the operator giving the density as
\be
\hat{\rho}(\rb) = \rho_0 + \delta\hat{\rho}(\rb)
\ee
where $\rho_0$ is a number, and one quadratizes the Hamiltonian
\be
\hat{H} = \int \hat{\psi}^\dagger h_0 \hat{\psi} 
+\frac{g}{2} \hat{\psi}^{\dagger 2} \hat{\psi}^2
\ee
in $\delta\hat{\rho}$ under the condition of weak density fluctuations,
that is of a small variance of $\hat{\rho}$:
\be
\mbox{Var}\hat{\rho}\equiv
\langle\hat{\rho}^2\rangle -\langle\hat{\rho}\rangle^2 \ll \rho^2
\label{eq:var_small}
\ee
where $\rho$ is the mean density.

\subsubsection{Problems with the general idea}\label{subsubsec:pbs}
Whereas the general idea works fine with a classical field model,
it is plagued by three major problems for the quantum field problem.

{\bf Problem 1}: the representation of the atomic interaction 
potential by $g$ times a Dirac distribution is perfectly fine in 1D,
but leads to a mathematically ill-defined quantum problem in 2D
and in 3D and gives rise to ultraviolet divergencies.

{\bf Problem 2}: The variance of $\hat{\rho}$ is infinite,
so that condition (\ref{eq:var_small}) is not satisfied.
The second moment of $\hat{\rho}$ can indeed be expressed
in terms of the $g_2$ function by using the bosonic commutation relations
for $\hat{\psi}$:
\be
\langle\hat{\rho}^2\rangle = \delta(\mathbf{0})\rho + g_2(\mathbf{0}) = +\infty.
\ee
Note that the function $g_2$ is bounded from above in any realistic model.
This divergence comes from `quantum fluctuations' of the field, that is from
the fact that $\hat{\rho}^2$ is not normally ordered with respect to $\hat{\psi}$.

We conclude that condition (\ref{eq:var_small}) is not the proper physical way of
defining the regime of weak density fluctuations. Everything becomes clear
if one considers the statistics of the number of particles in a {\it finite}
volume of space, e.g.\ a box of length $l$ and volume $l^d$.
The operator giving the number of particles in the box $B$ is
\be
\hat{n} = \int_B d^d\rb\, \hat{\psi}^\dagger(\rb)\hat{\psi}(\rb).
\ee
This operator has now finite relative mean squared fluctuations:
\be
\frac{\mbox{Var}\,\hat{n}}{\langle \hat{n}\rangle^2} =
\frac{1}{\langle\hat{n}\rangle} +
\int_B \frac{d^d\rb}{l^d}\int_B \frac{d^d\rpb}{l^d} 
\left[ \frac{g_2(\rb-\rpb)}{\rho^2}-1
\right].
\label{eq:var_n}
\ee
The first term on the right hand side of this expression is the quantum term leading
to the infinite variance of $\hat{\rho}$: it diverges in the limit $l\rightarrow 0$.
In the treatment to follow, we will consider a size of the box large enough
so that 
\be 
\langle\hat{n}\rangle=\rho l^d \gg 1.
\ee 
In the regime of weak density fluctuations 
the second term in the right hand side of (\ref{eq:var_n}) is also small.
If one can find a length $l$ smaller than the spatial scale of variation
of $g_2$, but still satisfying $\rho l^d\gg 1$, the regime of weak density fluctuations
corresponds to
\be
|\frac{g_2(\mathbf{0})}{\rho^2}-1| \ll 1.
\ee

{\bf Problem 3}: there exists no hermitian phase operator satisfying (\ref{eq:conj}).
We produce a simple proof {\it ad absurdum} \cite{other_proofs}. Imagine that there exists
a Hermitian operator satisfying the commutation relation (\ref{eq:conj}).
Then the following operator $\hat{T}(\alpha)$ is unitary:
\be
\hat{T}(\alpha) = e^{-i\alpha\hat{\theta}(\rzb)}
\ee
where $\alpha$ is real number and $\rzb$ is a fixed point in space.
Let us calculate the corresponding unitary transform of $\hat{n}$:
\be
\hat{n}(\alpha) \equiv \hat{T}^\dagger(\alpha) \hat{n} \hat{T}(\alpha).
\ee
Assume that the point $\mathbf{r}_0$ is inside the box $B$. Then the commutation 
relation (\ref{eq:conj}) implies
\be
[\hat{\theta}(\rzb),\hat{n}] = -i.
\ee
As a consequence, $\hat{n}(\alpha)$ obeys the differential equation
\be
\frac{d}{d\alpha}\hat{n}(\alpha) = e^{i\alpha\hat{\theta}(\rzb)}
i[\hat{\theta}(\rzb),\hat{n}]e^{-i\alpha\hat{\theta}(\rzb)} = 1
\ee
with the `initial' condition $\hat{n}(0)=\hat{n}$. This finally leads to
\be
\hat{n}(\alpha) = \hat{n} + \alpha.
\ee
We then have the following disaster:
if $|\psi\rangle$ is an eigenvector of $\hat{n}$ with eigenvalue $n$,
$\hat{T}(\alpha)|\psi\rangle$ is an eigenvector of $\hat{n}$ with eigenvalue $n+\alpha$.
As $\alpha$ can be any real number, this contradicts two well established
properties of the spectrum of $\hat{n}$, the discreteness and the positivity.

\begin{figure}
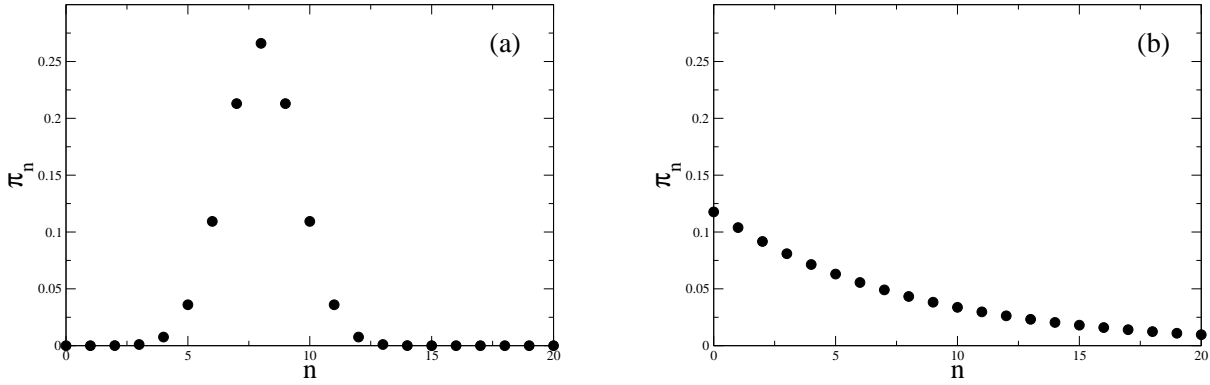

\centerline{\includegraphics[height=5cm,clip]{pia.eps}
\hspace{1cm}
\includegraphics[height=5cm,clip]{pib.eps}}
\caption{Two possible shapes for the probability distribution $\pi_n$ of the
number $n$ of particles in a box $B$ of volume $l^d$. (a) For a quasi-condensate
with an appropriate choice of $l$: the most probable value is $>0$, $\pi_n$ is approximately
Gaussian with a standard deviation smaller than the mean value.
(b) For an ideal Bose gas in the absence of Bose condensate: the most probable value
is $n=0$, $\pi_n$ is essential an exponential function of $n$. The mean value of $n$ is
8 in both examples.
\label{fig:pi_n}}
\end{figure}

Violation of discreteness is somehow attenuated by the fact that in practical
physical calculations, only integer values of $\alpha$ appear: in the Hamiltonian
and in the $g_1$ function for example, one has only $\alpha=0,+1$ or $-1$.
Violation of positivity is in general a real problem. It becomes in practice a
minor problem if the physical state of the system is such that
\be
\langle \hat{n}\rangle \gg 1 \ \ \ \mbox{and} \ \ \ \mbox{Var}\,\hat{n} \ll
\langle \hat{n}\rangle^2
\ee
and such that the probability of having $0$ or $1$ particle in the box $B$
is absolutely negligible.
In other words the use of the commutation relation (\ref{eq:conj}) is an
acceptable approximation when the probability distribution $\pi_n$ of the number
of particles in the box $B$ has a shape like in figure \ref{fig:pi_n}a, which is
typical of a quasi-condensate,
but not when $\pi_n$ has a shape like in figure \ref{fig:pi_n}b, which is typical
of the non-condensed ideal Bose gas.

\subsubsection{A solution to these three problems}
The introduction of a finite size box is a key ingredient of the previous
subsection \ref{subsubsec:pbs}. We therefore discretize the real space on a grid
with a step $l$, a square grid in 2D and a cubic grid in 3D.
The position $\rb$ now refers to the coordinates of the center of each cell.
The field operator $\hat{\psi}(\rb)$ annihilates a particle in the cell
of center $\rb$ and is normalized in such a way that
\be
[\hat{\psi}(\rb),\hat{\psi}^\dagger(\rpb)] =
\frac{\delta_{\rb,\rpb}}{l^d}
\ee
to recover the usual continuous quantum field theory for $l\rightarrow 0$.
We use periodic boundary conditions with lengths that are integer multiple of $l$.

{\bf Solution to problem 1}: the interaction potential is modeled by a discrete
delta potential
\be
V(\rb-\rpb) = \frac{g_0}{l^d} \delta_{\rb,\rpb}
\ee
where the choice of the so-called bare coupling constant $g_0$ is discussed
later. Note that, in 2D and 3D, $g_0$ strongly depends on $l$ in the limit
$l\rightarrow 0$, which is a signature of the pathology of a Dirac delta potential
in a continuous 2D or 3D space.

The grand canonical model Hamiltonian in second quantized form is then
\be
\hat{H} = \sum_{\rb} l^d\left[-\frac{\hbar^2}{2m}\hat{\psi}^\dagger(\rb)
\Delta\hat{\psi}(\rb) + (U(\rb)-\mu)\hat{\psi}^\dagger(\rb)\hat{\psi}(\rb)
+\frac{g_0}{2}\hat{\psi}^{\dagger 2}(\rb)\hat{\psi}^2(\rb)
\right]
\label{eq:hamil}
\ee
where we have introduced the discrete Laplacian $\Delta$. 
In the initial stage of the Bogoliubov approach to come, it is convenient to represent this
discrete Laplacian by
\be
\Delta f(\rb) = \sum_{j\in d\ {\rm directions\ of\ space}}
\frac{f(\rb+l\mathbf{e}_j)+f(\rb-l\mathbf{e}_j)-2f(\rb)}{l^2}.
\label{eq:3pt}
\ee
In this case our Hamiltonian becomes fully equivalent to the Bose Hubbard model
with an on-site interaction $U=g_0/l^d$ and a tunneling amplitude
$t=-\hbar^2/2ml^2$. The price to pay is that the dispersion relation of the
kinetic energy term is 
\be
\epsilon_{\kb}^{\mathrm{kin}} = \frac{\hbar^2}{ml^2}\sum_{j=1}^{d}
\left(1-\cos k_j l\right)
\label{eq:kincos}
\ee
which coincides with the correct parabola only in the range $kl\ll 1$.
In the final use of the results derived by the Bogoliubov approach, 
in particular
in a numerical treatment for the spatially inhomogeneous case, it is therefore 
more accurate to define $\Delta$ in Fourier space, simply by the requirement 
that the plane wave
of wavector $\mathbf{k}$ is an eigenstate with eigenvalue $\hbar^2 k^2/2m$,
each component $k_j$ of the wavevector being restricted to the first Brillouin
zone $[-\pi/l,\pi/l[$.

{\bf Solution to problem 2}: the variance of the density is finite.
In the discrete model, the operator giving the density is
$\hat{\rho}(\rb) = \hat{\psi}^\dagger(\rb)\hat{\psi}(\rb)$ and has the
physical meaning of being equal to $l^{-d} \hat{n}$ where $\hat{n}$ is the 
operator giving the number of particles in the considered cell.
The same phenomenon as in \S\ref{subsubsec:pbs} takes place, the variance
of $\hat{\rho}$ is finite:
\be
\mbox{Var}\,\hat{\rho} = \frac{\rho}{l^d} + 
[\langle\hat{\psi}^{\dagger 2}\hat{\psi}^2\rangle-\rho^2]
\label{eq:var_dis}
\ee
where $\rho$ is the mean density in the considered cell. We assume that both
terms in the right hand side of (\ref{eq:var_dis}) are much smaller
than $\rho^2$:
\bea
\rho l^d &\gg & 1 \\
|\langle\hat{\psi}^{\dagger 2}\hat{\psi}^2\rangle-\rho^2| \ll \rho^2.
\eea
The first condition ensures that the cell has a large occupation, and the
second one ensures that the density fluctuations are weak.

{\bf Solution to problem 3}: approximate construction of the phase operator.
Following \cite{Girardeau,Dum} one can introduce the exact writing
\be
\hat{\psi} \equiv \hat{A} \sqrt{\hat{\rho}}
\ee
where $\hat{A}$ decreases the number of particles in the cell by
one with an amplitude one rather than $\sqrt{n}$:
\bea
\hat{A} |n\rangle &=& | n-1\rangle \ \ \ \mbox{if} \ \ n>0 \\
                     &=& 0  \ \ \ \mbox{otherwise}.
\eea
The following identities are exact:
\bea
\hat{A} \hat{A}^\dagger &=& 1\\
\hat{A}^\dagger \hat {A} &=& 1 - |0\rangle\langle 0|
\eea
where $|0\rangle$ is the vacuum state of the considered cell.
This reveals that $\hat{A}$ is not unitary. However if the probability of having
no particle in the cell is truly negligible in the state on which 
$\hat{A}$ acts, the overlap of this state with $|0\rangle\langle 0|$ can be neglected
and one can assume that $\hat{A}$ is unitary:
\be
\hat{A}^\dagger \hat {A} \simeq \hat{A} \hat{A}^\dagger =1.
\ee
One then can set
\be
\hat {A} = e^{i\hat{\theta}}
\ee
where $\hat{\theta}$ is Hermitian. The commutation relation (\ref{eq:conj})
is then an acceptable approximation that ensures e.g. that
\be
\hat{A}^\dagger\hat{\rho}\hat {A} \simeq \hat{\rho} - \frac{1}{l^d}.
\ee
We note that there exists rigorous definitions of the phase operator \cite{phase}, but
the resulting operator does not satisfy (\ref{eq:conj}), which makes explicit
calculations more difficult.

\subsubsection{How to choose the grid spacing $l$ ?}
A first condition is that $l$ is large enough so that $\rho l^d\gg 1$. 
Note that the model Hamiltonian
(\ref{eq:hamil}) can be used outside this regime, this condition being useful
only for the Bogoliubov approximation to come.

A second condition is that $l$ is small enough so that the energy cut-off
$\sim \hbar^2/m l^2$ introduced by the grid does not change the physics.
This obviously requires that the energy cut-off is larger than $k_B T$
and than the chemical potential $\mu$. Equivalently
\be
l < \lambda, \xi
\ee
where $\lambda$ is the thermal de Broglie wavelength 
and $\xi$ is the healing length defined in (\ref{eq:def_xi}).

These three conditions are compatible only in the degenerate 
(\ref{eq:degen}) and weakly interacting 
(\ref{eq:weak}) regime to which our Bogoliubov approach is therefore restricted.

\subsubsection{How to choose the coupling constant $g_0$ ?}

The idea is the following. Consider the scattering of a plane wave on the
discrete $\delta$ potential on the grid, for a finite grid spacing $l$
but of course in the case of an infinite quantization volume. Calculate the
corresponding scattering amplitude and compare it to the exact amplitude
for the true interaction potential in continuous space.
Adjust $g_0$ to have the same scattering amplitude in the low energy 
domain.

Let us calculate the $T$ matrix for two interacting particles
in the discrete model. We take the version of the discrete
Laplacian giving the correct parabolic spectrum, as discussed after 
(\ref{eq:kincos}).
The center of mass motion can be separated from the relative motion
so we consider the reduced Hamiltonian for the relative motion of the two
particles on the grid:
\be
H = \frac{p^2}{m} + V
\ee
with $V=g_0\delta_{\rb,\mathbf{0}}/l^d$. The eigenstates of the kinetic
energy are plane waves with a wavevector $\mathbf{k}$, with a wavefunction
\be
\langle\rb|\kb\rangle = e^{i\kb\cdot\rb}.
\ee
As each component $r_j$ is an integer multiple of $l$, the component $k_j$
has a meaning modulo $2\pi/l$ and can be restricted to the first Brillouin
zone $[-\pi/l,\pi/l[$. We normalize the localized state vector $|\rb\rangle$
so as to recover the continuous theory for $l\rightarrow 0$:
\be
\langle\rb |\rpb\rangle  = \frac{\delta_{\rb,\rpb}}{l^d}.
\label{eq:norma}
\ee
In this case the potential $V$ can be written as
\be
V = g_0 |\rb=\mathbf{0}\rangle\langle \rb=\mathbf{0}|
\ee
and the following closure relation holds:
\be
\int_D \frac{d^d\kb}{(2\pi)^d} |\kb\rangle\langle\kb| = 1
\label{eq:ferm}
\ee
where $D=[-\pi/l,\pi/l[^d$ is the first Brillouin zone of the reciprocal lattice.

The $T$ matrix at a given energy $E$ is defined by \cite{scattering}:
\be
T(E+i\eta) = V + V G(E+i\eta) V
\ee
where $\eta\rightarrow 0^+$ at the end of the calculation. $G$
is the resolvent of the full Hamiltonian:
\be
G(z) = \frac{1}{z-H}
\ee
defined for any complex number $z$ not belonging to the spectrum of $H$.
The particular form of $V$ leads to the simple expression for the matrix
element of $T$ in between two arbitrary plane waves \cite{arbitrary}:
\be
\langle\kb|T_{\rm grid}(E+i\eta)|\kpb\rangle = g_0 + g_0^2 \langle\rb=
\mathbf{0}| G(E+i\eta)|\rb=\mathbf{0}\rangle
\ee
so that only the matrix element of the resolvent in the state localized in the cell
$\rb=\mathbf{0}$ of the lattice matters!
This matrix element is immediately deduced from the recursion relation
\be
G(z) = G_0(z) + G_0(z) V G(z)
\ee
where $G_0$ is the resolvent of the kinetic energy operator. We finally obtain
\be
\langle\kb|T_{\rm grid}(E+i\eta)|\kpb\rangle = \frac{g_0}{1-g_0 I(E)}
\ee
where 
\be
I(E) =  \langle\rb=\mathbf{0}| G_0(E+i\eta)|\rb=\mathbf{0}\rangle
= \lim_{\eta\rightarrow 0^+} \int_D \frac{d^d\kb}{(2\pi)^d} \frac{1}{E+i\eta -\hbar^2k^2/m}.
\ee
Note that, due to the discrete delta nature of the interaction potential, the matrix
element of $T_{\rm grid}$ depends only on the energy, not on the wavevectors: the operator $T_{\rm grid}$ is
actually proportional to $|\rb=\mathbf{0}\rangle \langle\rb=\mathbf{0}|$.

To calculate $I(E)$ one may use the following identity for distributions:
\be
\lim_{\eta\rightarrow 0^+} \frac{1}{X+i\eta} = {\cal PP}\frac{1}{X} -i\pi
\delta(X)
\ee
where ${\cal PP}$ is the principal part. We restrict to positive energies $E$
and we set $E=\hbar^2q^2/m$. We are interested in the low energy limit $E\ll\hbar^2/m l^2$
that is $q \ll \pi/l$. Why ? In the degenerate and weakly interacting regime
the maximal value of 
$q$ is $1/\xi$ or $1/\lambda$, which is much smaller than $1/l$.

The imaginary part of $I(E)$ is easy to calculate when $q< \pi/l$, in which case
the support of the $\delta$ distribution is inside the Brillouin zone $D$ and $\mbox{Im}\, I(E) $
is proportional to the density of states:
\be
\begin{array}{ccccc}
\mbox{Im}\, I(E)  & =& \displaystyle -\frac{m}{4\pi\hbar^2} q  &  \mbox{in 3D}  & \\
               &&&& \\
                 & =& \displaystyle -\frac{m}{4\hbar^2}       &  \mbox{in 2D} 
& \ \ \ \ \mathrm{for}\ \ q<\pi/l \\
               &&&& \\
                  & =& \displaystyle -\frac{m}{2\hbar^2}\frac{1}{q}  &\mbox{in 1D} &
\end{array}
\label{eq:im3d}
\ee

The calculation of the real part of $I(E)$ is more involved:
\be
\mbox{Re}\, I(E) = \frac{m}{\hbar^2}
\int_D \frac{d^d\kb}{(2\pi)^d} {\cal PP} \frac{1}{q^2-k^2}
\ee
so we shall restrict to the low energy limit. In 3D, the real part has a finite
limit for $q\rightarrow 0$: replacing $q$ by zero, one obtains
\bea
\mbox{Re}\, I(E=0) &=& \frac{m}{\hbar^2} \int_D \frac{d^3\kb}{(2\pi)^3} \frac{1}{k^2} \\
&=& \frac{m}{\hbar^2 l} \times 0.194 \ldots
\label{eq:re3d}
\eea
The situation is dramatically different in 2D, where one gets a divergent expression
if one replaces $q$ by zero. The trick is then to split the integration domain
$D$ in the disk of radius $\pi/l$, over which the integral can be calculated exactly,
and in the complementary domain, where $q^2$ is a small perturbation of $k^2$
\cite{Mora}:
\be
\mbox{Re}\, I(E) = \frac{m}{2\pi\hbar^2}
\left[ \ln(ql/\pi)-C + O((ql)^2)
\right]
\ee
with the constant
\be
C= \ln 2 -\frac{2}{\pi} G = 0.110025 \ldots
\ee
involving Catalan's constant $G=0.915965594\ldots$.
In 1D the exact calculation can be done:
\be
\mbox{Re}\, I(E) = \frac{m}{2\pi\hbar^2 q}\ln\left|
\frac{\pi+ql}{\pi-ql}\right| \sim_{ql\rightarrow 0} \frac{ml}{\pi^2\hbar^2}.
\ee

The last step is to compare with the scattering amplitude of the true interaction
potential in the continuous case.
In 3D a low energy approximation can be used for the true $T$ matrix:
\be
\langle\kb|T_{\rm true}(E+i\eta)|\kpb\rangle \simeq 
\frac{4\pi\hbar^2/m}{\frac{1}{a}+iq},
\ee
where $a$ is the $s$-wave scattering length,
provided that $k r_e, k' r_e, q r_e \ll 1$, where $r_e$ is the effective range of the true interaction
potential. The maximal value of $q,k, k'$ that one may expect to appear in the thermal state
of the gas in the degenerate regime is the inverse of the mean interparticle separation $\rho^{1/3}$,
whatever the strength of the interactions: one has
to check that $\rho r_e^3 \ll 1$, which is the case in present experiments even
close to a Feshbach resonance \cite{Combescot}. In the weakly interacting regime, 
the situation is even more favorable as the typical $q,k,k'$ is at most
$1/\xi$ or $1/\lambda$, much smaller than $\rho^{1/3}$.

Using the expressions (\ref{eq:re3d}) and (\ref{eq:im3d}) for $I(E)$ and identifying
$T_{\rm grid}$ with $T_{\rm true}$ leads to the identity
\be
\frac{1}{g_0} = \frac{m}{4\pi\hbar^2 a} - \frac{m}{\hbar^2} \int_D \frac{d^3\kb}{(2\pi)^3} 
\frac{1}{k^2}
\label{eq:g03d}
\ee
that is
\be
g_0 = \frac{4\pi\hbar^2 a/m}{1-K_3 a/l} \ \ \ \mbox{with} \ \ 
K_3=2.442\, 749 \ldots
\ee
Note the exact cancellation of the imaginary parts of the denominators
of $T_{\rm grid}$ and $T_{\rm true}$.
In the considered weakly interacting regime, condition (\ref{eq:weak}) implies
$\sqrt{\rho a^3}\ll 1$; as $\rho l^3\gg 1$ one finds that $a\ll l$ so that $g_0$
is close to the usual coupling constant $g=4\pi\hbar^2 a/m$. Close, but not identical, and
this plays an important role in suppressing potential ultraviolet divergences.

In 2D, the low energy approximation for the `true' $T$ matrix is
\be
\langle\kb|T_{\rm true}(E+i\eta)|\kpb\rangle \simeq 
\frac{-2\pi\hbar^2/m}{\ln(a_2 q/2)+\gamma -i\pi/2}
\ee
where $a_2$ is the 2D scattering length and $\gamma=0.57721\ldots$
is Euler's constant, see the lecture of Gora Shlyapnikov in this volume.
In identifying with the low energy expression of $T_{\rm grid}$, one finds again that
the imaginary parts of the denominators of $T_{\rm grid}$ and $T_{\rm true}$
exactly cancel and one is left with:
\be
g_0 =\frac{2\pi\hbar^2}{m}\frac{1}{\ln(K_2 l/a_2)}
\label{eq:g02d}
\ee
with a numerical constant 
\be
K_2 = \frac{1}{\pi} e^{-\gamma + 2G/\pi}\simeq \frac{1}{\pi}
\ee
where we used the amusing fact that $-\gamma + 2G/\pi=5.91\ldots 10^{-3} \ll 1$.

In 1D one may model the continuous interaction potential by $g_1\delta(z_1-z_2)$,
under validity conditions discussed in \cite{Olshanii,Olshanii2}, in which case
the $T$ matrix is
\be
\langle k|T_{\rm true}(E+i\eta)| k'\rangle \simeq
\frac{\hbar^2/m}{a_1+i/2 q}
\label{eq:T1d}
\ee
where the 1D scattering length is such that
\be
g_1 = \frac{\hbar^2}{m a_1}.
\ee
The identification of $T_{\rm grid}$ and $T_{\rm true}$  leads to
\be
g_0 = \frac{g_1}{1+l/(a_1\pi^2)}.
\ee
In the weakly interacting regime, $g_0$ is very close to $g_1$:
$l\ll \xi$, so that $l/a_1 \ll \xi/a_1\propto 1/\sqrt{\rho a_1}\propto 1/\rho \xi \ll 1$.
Contrarily to the 3D case, the small difference between $g_0$ and $g_1$
does not play a significant role. 

\subsection{Perturbative expansion of the model Hamiltonian}
In a regime where each cell has a negligible probability of being empty,
we may introduce the phase operator having approximately the commutation
relation (\ref{eq:conj}) with the density and write the Hamiltonian as
\bea
\hat{H} &=& \hat{H}_{\rm kin} + \hat{H}_{\rm pot} \\
\hat{H}_{\rm pot} &=& \sum_{\rb} l^d \hat{\rho}(\rb)
\left[ U(\rb) -\mu +\frac{g_0}{2} \left(\hat{\rho}(\rb) -\frac{1}{l^d} \right) \right] \\
\hat{H}_{\rm kin} &\simeq& -\frac{\hbar^2}{2m l^2} \sum_{\rb}
l^d \sum_{j=1}^{d} \left\{\left[\hat{\rho}^{1/2} e^{i(\hat{\theta}_{+j}-\hat{\theta})}
\hat{\rho}_{+j}^{1/2} +\mbox{h.c.} \right] - 2 \hat{\rho} \right\}
\eea
where we have introduced the notation
\bea
f_{+j} &=& f(\rb+l\mathbf{e}_j) \\
f_{-j} &=& f(\rb-l\mathbf{e}_j)
\eea
for each direction of space.

\subsubsection{Two small parameters}
The first small parameter expresses the weakness of the density fluctuations: one splits
\be
\hat{\rho}(\rb) = \rho_0(\rb) + \delta\hat{\rho}(\rb)
\ee
where $\rho_0$, the zeroth order approximation to the density, corresponds
to a pure quasi-condensate. The small parameter is then
\be
\epsilon_1 = \frac{|\delta\hat{\rho}|}{\rho_0} \ll 1.
\ee

The second small parameter expresses the smallness of the phase variation in between
neighboring sites of the grid:
\be
\epsilon_2 = | l\, \mathbf{grad}\,\hat{\theta} | \ll 1
\ee
where $\mathbf{grad}$ is the discrete gradient on the lattice.

It can be checked at the end of all calculations that by an appropriate choice
of $l$ one can achieve
\be
\epsilon_1 \sim \epsilon_2 \sim \frac{1}{\sqrt{\rho_0 l^d}}.
\ee
One may note that this value corresponds to the `quantum' fluctuation part,
that is the first term in the right-hand side of (\ref{eq:var_dis}).
An important  consequence is that the two small parameters may be considered
as a single small parameter, in the expansion of the Hamiltonian.

\subsubsection{Quadratisation of the Hamiltonian}

Expansion of $\hat{H}$ in powers of $\epsilon_1$ and $\epsilon_2$
is performed from the series expansions
\bea
\hat{\rho}^{1/2} &=& \rho_0^{1/2} +\frac{\delta\hat{\rho}}{2\rho_0^{1/2}}
-\frac{\delta\hat{\rho}^2}{8\rho_0^{3/2}}+
\frac{\delta\hat{\rho}^3}{16\rho_0^{5/2}}+\ldots \\
e^{i(\hat{\theta}_{+j}-\hat{\theta})} &=& 
1 + i(\hat{\theta}_{+j}-\hat{\theta}) 
-\frac{1}{2} (\hat{\theta}_{+j}-\hat{\theta})^2
+\ldots
\eea

To zeroth order in the small parameters $\epsilon_1$ and $\epsilon_2$,
density fluctuations are neglected, $\hat{\rho}$ being approximated
by $\rho_0$, and the spatial variation of the phase operator is also
neglected. This leads to the zeroth order approximation to the Hamiltonian,
which is the following c-number:
\be
H_0 = \sum_{\rb} l^d \rho_0^{1/2} \left[
-\frac{\hbar^2}{2m} \Delta +U(\rb) +\frac{1}{2} g_0 \rho_0-\mu\right]
\rho_0^{1/2}.
\ee
In 3D, 
this is the equivalent on the lattice of the Gross-Pitaevskii energy functional,
with the difference that the bare coupling constant $g_0$ appears, rather than $g$.
$\rho_0$ is then naturally obtained by minimization of $H_0$, so that it 
obeys:
\be
\left[-\frac{\hbar^2}{2m} \Delta +U(\rb) + g_0 \rho_0-\mu\right]\rho_0^{1/2} = 0.
\label{eq:gptype}
\ee
This equation naturally defines $\rho_0$ as a function of $\mu$.
However it will reveal more convenient to parameterize the theory
in terms of $N_0$, i.e. the total number of particles stored
in the density profile $\rho_0$:
\be
N_0 = \sum_{\rb} l^d \rho_0(\rb).
\ee
We will therefore consider $\mu$ and $\rho_0$ as functions 
of $N_0$:
\bea
\mu &=& \mu_0(N_0) 
\label{eq:mu0}\\
\rho_0(\rb) &=& \rho_0(\rb;N_0).
\eea

As a consequence of
Eq.(\ref{eq:gptype}), 
one finds that the first order approximation $\hat{H}_1$
to the Hamiltonian exactly vanishes. The first non-trivial contribution
is therefore a quadratic one:
\bea
\hat{H}_2 & = &  E_2[\rho_0] + \sum_{\rb} l^d \left[
- \frac{\hbar^2}{2 m} \frac{\delta\hat{\rho}}{2 \sqrt{\rho_0}} \Delta
\left(\frac{\delta\hat{\rho}}{2 \sqrt{\rho_0}}\right) +
\frac{\hbar^2 \delta\hat{\rho}^2}{8 m \rho_0^{3/2}}
\Delta \sqrt{\rho_0} + \frac{g_0}{2} \delta\hat{\rho}^2\right. \\  & + &
\left.\frac{\hbar^2}{2 m}  \sum_{j}
\sqrt{\rho_0(\rb  ) \rho_0(\rb   + l  {\bf e_j})}
\, \frac{(\hat{\theta}(\rb+l{\bf e_j})-\hat{\theta}(\rb))^2}{l^2} \right] \\
\eea
where the c-number energy functional $E_2[\rho_0]$ is given by
\be
E_2[\rho_0] = -\frac{g_0}{2} \sum_{\rb}\rho_0
-\frac{\hbar^2}{4m l^2} \sum_{\rb} \sum_{j=1}^{d}  
\left[
\left(\frac{\rho_{0,+j}}{\rho_0}\right)^{1/2}+
\left(\frac{\rho_0}{\rho_{0,+j}}\right)^{1/2}
\right].
\ee
Remarkably, one finds that the Hamiltonian $\hat{H}_2$ is equivalent
to the Hamiltonian of the $U(1)$-symmetry breaking Bogoliubov approach 
for the usual case of true condensates: one introduces the field
\be
\hat{B}(\rb) = \frac{\delta\hat{\rho}(\rb)}{2\rho_0^{1/2}(\rb)}
+i \rho_0^{1/2}(\rb) \hat{\theta}(\rb).
\ee
One can then check that it has bosonic commutation relations
\be
[\hat{B}(\rb),\hat{B}^\dagger(\rpb)] = \frac{\delta_{\rb,\rpb}}{l^d}.
\ee
After some algebra, using the fact that $\rho_0^{1/2}$ solves the
Gross-Pitaevskii type equation (\ref{eq:gptype}), one obtains
the following exact rewriting of $\hat{H}_2$:
\be
\hat{H}_2 = l^d \sum_{\rb}
\hat{B}^\dagger \left(-\frac{\hbar^2}{2m} \Delta + U + g_0 \rho_0
-\mu \right)\hat{B}
+g_0\rho_0 \left[\hat{B}^\dagger \hat{B} +
\frac{1}{2} (\hat{B}^2+\hat{B}^{\dagger 2})
\right]
\ee
which has exactly the structure of the usual Bogoliubov Hamiltonian.
Remarkably, the c-number energy $E_2$ is exactly compensated by the
contribution of commutators of $\delta\hat{\rho}$ with $\hat{\theta}$.
One then reuses the standard diagonalization of the Bogoliubov
Hamiltonian:
\be
\hat{B}(\rb) = -i \hat{Q}\rho_0^{1/2}(\rb) +
\hat{P} \partial_{N_0} \rho_0^{1/2}(\rb)
+\sum_s \hat{b}_s u_s(\rb) + \hat{b}^\dagger_s v_s^*(\rb)
\ee
as explained in \cite{Blaizot,Lewenstein,Dum}.
Here the sum $s$ is taken over the regular Bogoliubov eigenmodes
$(u_s,v_s)$ 
\be
\left(
\begin{array}{cc}
-\frac{\hbar^2}{2m}\Delta + U - \mu + 2 g_0 \rho_0  & g_0 \rho_0 \\ 
- g_0\rho_0 & 
- \left( -\frac{\hbar^2}{2m}\Delta + U - \mu + 2 g_0 \rho_0 \right)
\end{array} 
\right)
\left(\begin{array}{c} u_s \\ v_s  \end{array}\right)=
\epsilon_s \left(\begin{array}{c} u_s \\ v_s  \end{array}\right)
\ee
normalizable as
\be
\sum_{\rb} l^d \left[|u_s(\rb)|^2 - |v_s(\rb)|^2 \right] = 1.
\ee
Thermodynamic stability, or equivalently the fact that
$\rho_0$ is a minimum of $H_0$,  imposes that the corresponding eigenenergies
$\epsilon_s$ are positive
\cite{Houches99}. The operators $\hat{b}_s$ obey the usual bosonic
commutation relations $[\hat{b}_s,\hat{b}^\dagger_{s'}]=\delta_{s,s'}$.
The operators $\hat{Q}$ and $\hat{P}$ 
commute with $\hat{b}_s,\hat{b}_s^\dagger$ and 
are conjugate quantum variables,
\be
[\hat{P},\hat{Q}]=-i.
\ee
$\hat{Q}$ is a collective coordinate representing the quantum phase
of the field, and, as shown in \cite{Mora}, $\hat{P}$ gives the fluctuations
of the total number of particles away from the total number of particles 
$N_0$ contained in the density profile $\rho_0$:
\be
\hat{P}=l^d \sum_{\rb} [\hat{\rho}(\rb)-\rho_0(\rb)] = \hat{N}-N_0.
\ee
We note that our treatment is in the grand canonical
ensemble but does not break the $U(1)$-symmetry of the
Hamiltonian: the emergence of the operators $\hat{P}$ and $\hat{Q}$
is therefore a consequence of a non-fixed value of the total number
of particles, rather than of symmetry breaking.

Replacing $\hat{B}$ by the Bogoliubov modal decomposition 
finally leads to a modal expansion for the operators
giving the density and the phase:
\bea
\hat{\theta}(\rb) &=& - \hat{Q} +
\sum_s \theta_s (\rb) \, \hat{b}_s + \theta_s^*(\rb) 
\hat{b}_s^{\dagger} \\
\delta\hat{\rho}(\rb) &=& 
\hat{P}\partial_{N_0}\rho_0(\rb) +
\sum_s \delta\rho_s(\rb) \hat{b}_s + \delta\rho_s^*(\rb)
\hat{b}_s^\dagger 
\eea
where
\bea
\theta_s (\rb) &=& \frac {u_s(\rb)-v_s(\rb)}{2i\rho_0^{1/2}(\rb)}\\
\delta\rho_s(\rb) &=& \rho_0^{1/2}(\rb)[u_s(\rb)+v_s(\rb)].
\eea
It also gives the normal form of the Hamiltonian $\hat{H}_2$,
from which thermal averages can be evaluated easily:
\be
\hat{H}_2 = -\frac{1}{2} N_0 \frac{d\mu_0}{dN_0}
-\sum_s \epsilon_s \langle v_s|v_s\rangle
+\frac{1}{2} \frac{d\mu_0}{dN_0} \hat{P}^2 +\sum_s 
\epsilon_s \hat{b}_s^\dagger \hat{b}_s
\ee
where the function $\mu_0$ is defined in Eq.(\ref{eq:mu0}) 
\cite{premier_terme}.

\subsubsection{Why cubic terms in the Hamiltonian are required}

One could believe that the knowledge of $\hat{H}_2$ is sufficiently 
to calculate to a given order $O(\epsilon_{1,2}^2)$ the correction 
for any observable to the pure quasi-condensate approximation.
However this is not true,  the mean density being an obvious
counter-example: in a thermal state density operator with the Hamiltonian
$\hat{H}_2$, the mean value of $\delta\hat{\rho}$ vanishes so that, for a
given chemical potential, there is no correction to the mean density 
as compared to the pure quasi-condensate assumption.

A similar phenomenon takes place in the quantum mechanical
problem of a single particle strongly confined
in a non-harmonic trapping 
potential $V(x)$ in 1D: a series expansion of the potential is performed
around its minimum,
\be
V(x)= V_0 + \frac{1}{2} V''(x_0) (x-x_0)^2+\frac{1}{6}V^{(3)}(x-x_0)^3+
\ldots
\label{eq:serie}
\ee
where $x$ is the particle coordinate. To zeroth order in the expansion,
the ground state energy of the particle is $V_0$ and its position
is $x_0$. The first correction
to the energy is given by the inclusion of the quadratic term of
Eq.(\ref{eq:serie}) and, when expressed in units of $V_0$,
it scales as $V''^{1/2}/V_0 \sim 1/(V_0^{1/2}x_0)$
where $x_0$ is assumed to be also a typical length scale of $V(x)$. 
The first correction
to the mean position of the particle is given by the inclusion of
the {\bf cubic} term in Eq.(\ref{eq:serie}) as a perturbation to the
harmonic potential and, when expressed in units of $x_0$, scales also
as $1/(V_0^{1/2}x_0)$!

Coming back to the quantum gas problem, one has to produce a third
order expansion of the model Hamiltonian and treat the resulting 
cubic Hamiltonian $\hat{H}_3$ as a perturbation to $\hat{H}_2$.
The derivation of $\hat{H}_3$ is detailed in an appendix of
\cite{Mora}, we give here only the result:
\bea
\hat{H}_3 &=& -\frac{g_0}{2}
\sum_{\rb}\delta\hat{\rho}
+\frac{\hbar^2}{4ml^2}
\sum_{\rb,j} l^d\,
(\hat{\theta}_{+j}-\hat{\theta})
\left(
\frac{\rho_{0,+j}^{1/2}}{\rho_0^{1/2}}\,\delta\hat{\rho}+
\frac{\rho_0^{1/2}}{\rho_{0,+j}^{1/2}}\,\delta\hat{\rho}_{+j} \right)
(\hat{\theta}_{+j}-\hat{\theta})
 \\
 &+&
\frac{\hbar^2}{8m} \sum_{\rb}
\frac{\delta\hat{\rho}}{\rho_0}
\left( \rho_0^{-1/2}\Delta\rho_0^{1/2}-\rho_0^{1/2}\Delta\rho_0^{-1/2} \right)
-\frac{\hbar^2}{16m}\sum_{\rb} l^D
\left[\frac{\delta\hat{\rho}^3}{\rho_0^{5/2}}\Delta\sqrt{\rho_0}
-\frac{\delta\hat{\rho}^2}{\rho_0^{3/2}}\Delta\left(\frac{\delta\hat{\rho}}{\sqrt{\rho_0}}\right)
\right].
\nonumber
\eea
There are then two main ways to calculate the correction to the
mean density due to $\hat{H}_3$. The first one relies on finite temperature
perturbation theory: to calculate expectation values in the density
operator $\exp[-\beta(\hat{H}_2 + \hat{H}_3)]$ to first order in $\hat{H}_3$,
one uses the imaginary time version of the time dependent first order
perturbation theory:
\be
e^{-\beta(\hat{H}_2 + \hat{H}_3)} =
e^{-\beta\hat{H}_2} - \int_0^{\beta}d\tau\, e^{-(\beta-\tau)\hat{H}_2}
\hat{H}_3 e^{-\tau \hat{H}_2} + \ldots
\label{eq:tdpt}
\ee
One is then back to the calculation of expectation values of operators
in the thermal state corresponding to $\hat{H}_2$, and Wick's theorem
can be applied.

The second way to include the effect of $\hat{H}_3$ perturbatively
is here simpler and was followed in \cite{Mora}. It
consists in writing the equation of motion of $\hat{\theta}$
in Heisenberg representation for the
Hamiltonian $\hat{H}_2 + \hat{H}_3$: with respect to the evolution governed
by $\hat{H}_2$, `new' terms appear. One then takes the expectation
value of this equation in the desired thermal state. The expectation
value of the `new' terms can be taken in the unperturbed thermal state,
as they originate from the perturbation $\hat{H}_3$. Furthermore
the expectation value of $\partial_t \hat{\theta}$ vanishes in steady
state, as proved in an appendix of \cite{Mora}, so that
\bea
0 &=& \left[ -\frac{\hbar^2}{2m}\Delta + U +3g_0\rho_0 -\mu \right]
\left(\frac{\langle\delta\hat{\rho}\rangle_3-\langle\hat{B}^\dagger
\hat{B}\rangle_{2}}{\rho_0^{1/2}} \right)
\nonumber \\
&+& g_0\rho_0^{1/2}\langle 4\hat{B}^\dagger\hat{B}+
\hat{B}^2+\hat{B}^{\dagger 2} \rangle_2-
2\langle \hat{P}^2\rangle_2\frac{d\mu_0}{dN_0}\partial_{N_0}\sqrt{\rho_0}
\label{eq:pour_rho3}
\eea
where the thermal average $\langle\ldots\rangle_2$
is taken with the unperturbed Hamiltonian $\hat{H}_2$ and
$\langle\ldots\rangle_3$ is taken with the perturbed Hamiltonian
$\hat{H}_2+\hat{H}_3$ to first order in $\hat{H}_3$.
The fact that $\rho_0$ is a minimum of $H_0$ imposes that the
differential operator in the first line of Eq.(\ref{eq:pour_rho3})
is strictly positive \cite{Dum,Houches99} so that this
equation determines $\langle\delta\hat{\rho}\rangle_3$ in a unique
way.

\subsection{Applications of the formalism}

We present here without derivation 
simple applications of the expansion of the Hamiltonian
performed in the previous subsection. For simplicity, we restrict
to the spatially homogeneous case of a gas with periodic boundary
conditions in a box of size $L$ along
each direction of space and we take the thermodynamic limit
$L\rightarrow +\infty$ for a fixed value of the chemical
potential.
The derivation of the formulas given below and their extension
to the case of a finite size, spatially inhomogeneous system can be
found in \cite{Mora}.

\subsubsection{Equation of state}
Thanks to Eq.(\ref{eq:pour_rho3}) we can relate the mean density
to the chemical potential to first order beyond the pure quasi-condensate
approximation. As derived in \cite{Mora}
\be
\frac{\mu}{g_0} = \rho +  \int_D \frac{d^d\kb}{(2\pi)^d} \left[
(U_k+V_k)^2 n_k + V_k (U_k + V_k) \right]
\label{eq:eqet}
\ee
where $n_k=1/[\exp(\beta\epsilon_k)-1]$ is the mean occupation
number of the Bogoliubov mode of wavevector $\kb$ and $U_k,V_k$
are the amplitudes of the Bogoliubov mode functions
$u_{\kb}(\rb)= U_k \exp(i\kb\cdot\rb)/L^{d/2}$
and $v_{\kb}(\rb)= V_k \exp(i\kb\cdot\rb)/L^{d/2}$:
\be
U_k + V_k = \frac{1}{U_k-V_k}=\left[
\frac{\hbar^2 k^2/2m} {2\mu +\hbar^2 k^2/2m}
\right]^{1/4}.
\ee
The corresponding Bogoliubov eigenenergy is
\be
\epsilon_k=\left[\frac{\hbar^2k^2}{2m}\left(\frac{\hbar^2k^2}{2m}+2\mu
\right) \right]^{1/2}.
\ee
In what follows we shall need the large $k$ dependence of the zero
temperature value of the integrand in Eq.(\ref{eq:eqet}):
\be
V_k (U_k + V_k) \simeq -\frac{m \mu}{\hbar^2 k^2}.
\label{eq:asymp}
\ee

Eq.(\ref{eq:eqet}) is not totally satisfactory yet as it depends
on the grid spacing $l$ of our lattice model, both through
the integration domain $D$ defined after Eq.(\ref{eq:ferm})
and through the $l$-dependence of the coupling constant $g_0$.
Let us check, for each value of the dimension $d$, that the $l$
dependence disappears in the limit $l\ll \xi,\lambda$.

In 1D this is clearly the case as both $g_0$ and the integral
in Eq.(\ref{eq:eqet}) have a finite limit when $l\rightarrow 0$.
One can then replace $g_0$ by the 1D coupling constant $g_1$,
and $D$ by $]-\infty,+\infty[$.

In 2D the integral in Eq.(\ref{eq:eqet}) diverges logarithmically
in the limit $l\rightarrow 0$, because of the $T=0$ contribution
to the integrand. We calculate the low-$l$ behavior of the integral
by splitting the integration domain $D$ in a disk of radius $\pi/l$,
over which the integral is performed in polar coordinates, and
a complementary domain, over which the integrand is approximated
by its asymptotic behavior Eq.(\ref{eq:asymp}).
We obtain
\be
-\int_D \frac{d^2 \kb}{(2\pi)^2} \, V_k (U_k + V_k) =
\frac{m\mu}{4\pi\hbar^2}
\left[\ln\left(\frac{\pi^2\hbar^2}{ml^2\mu}\right)-1 +2\ln 2 -4 G/\pi
+ O(l^{2}/\xi^2)
\right]
\ee
where $G$ is Catalan's constant. Remarkably this compensates the logarithmic
$l$ dependence in the value of $1/g_0$, see Eq.(\ref{eq:g02d}).
We arrive at the equation of state of the 2D gas:
\be
\rho = \frac{m \mu}{4 \pi \hbar^2} \ln \left( \frac{4 \hbar^2}{a_2^2 m \mu  
e^{2 \gamma +1}} \right) -  
\int\frac{d\kb}{(2 \pi)^2}  ( \bar{u_k} + \bar{v_k} )^2 n_k 
\ee
where $\gamma$ is Euler's constant and $a_2$ is the 2D scattering length.
This relation is identical to the result (20.45) obtained by the
functional integral method in \cite{Popov}.
It allows to show that the necessary condition $\rho\xi^2\gg 1$
for our treatment to apply imposes $\ln ( 1 / \rho a_2^2 ) \gg 4\pi$
at zero temperature.

In 3D, the integral in Eq.(\ref{eq:eqet}) also diverges.
Using the identity Eq.(\ref{eq:g03d}) satisfied by the bare coupling 
constant $g_0$ amounts to subtracting the high $k$ behavior
of the integrand, so that the $l\rightarrow 0$ limit may be taken to give
\be
\mu = \rho g +g \int \frac{d\kb}{(2 \pi)^3} \left(
(U_k+V_k)^2 n_k+V_k (
U_k + V_k ) + \frac{m \mu}{\hbar^2 k^2} \right).
\ee

In 3D our result coincides with the one of the usual Bogoliubov
approach for a condensate. What might be surprising at first sight
is that our results in 2D and in 1D also coincide with the ones
predicted by a blind application of the usual Bogoliubov method,
even if there is no true condensate! The same conclusion applies
for the ground state energy of a gas of $N$ particles, as shown in
\cite{Mora}, and as could be expected from the
equivalence of $\hat{H}_2$ with the usual Bogoliubov Hamiltonian. 
This fact that the usual Bogoliubov approach gives
the correct result was commonly used in the literature in 1D,
see e.g. \cite{Lieb}, and was justified by an asymptotic treatment
of the solution based on the Bethe ansatz in \cite{Gaudin}.
With our extended Bogoliubov method, we reach this conclusion in
a simpler, more transparent and more general way.

\subsubsection{Density fluctuations}
Our treatment relies on the assumption of weak density fluctuations.
We then have to check that the relative variance of the density fluctuations
is weak:
\be
\epsilon_1^2 = \frac{\langle \delta\hat{\rho}^2(\mathbf{0})\rangle_2}{\rho_0^2}
\stackrel{?}{\ll}1.
\ee
Following the discussion of \S\ref{subsubsec:pbs} 
we expect this variance
to be the sum of two contributions, one coming from the fact that 
$\delta\hat{\rho}^2$ is not a normally ordered operator, and a second
one involving the second order correlation function of the field,
\be
g_2(\rb) \equiv \langle \hat{\psi}^\dagger(\rb)\hat{\psi}^\dagger(\mathbf{0})
\hat{\psi}(\mathbf{0})\hat{\psi}(\rb)\rangle.
\label{eq:def_g2}
\ee

Explicit calculations are performed in \cite{Mora}.
In the thermodynamical limit one finds indeed that
\be
\frac{\langle \delta\hat{\rho}^2(\mathbf{0})\rangle_2}{\rho_0^2}=
\frac{1}{\rho_0 l^d} + 
\frac{\langle :\delta\hat{\rho}^2(\mathbf{0}):\rangle_2}{\rho_0^2}
\ee
where $:\, :$ is the standard notation to represent normal order
(with all the $\hat{\psi}^\dagger$ on the left and all the $\hat{\psi}$ 
on the right) and
\be
\frac{\langle :\delta\hat{\rho}^2(\mathbf{0}):\rangle_2}{\rho_0^2}=
\frac{2}{\rho_0}
\int_D \frac{d^d\kb}{(2\pi)^d} [(U_k+V_k)^2 n_k +
V_k (U_k +V_k)].
\label{eq:no}
\ee
To ensure the condition $\epsilon_1 \ll 1$, we require
that $\rho_0 l^d \gg 1$  and that the normal-ordered
contribution is small. The zero temperature contribution
to Eq.(\ref{eq:no}) is found in \cite{Mora} to be always
smaller than $1/\rho_0 l^d$ as soon as $l<\xi$, see the table below
in the column `quantum term'.
The same conclusion holds for the thermal contribution
at temperatures $k_B T < \mu$, cf.\ table.
At temperatures $k_B T > \mu$, the estimates produced in \cite{Mora} are
summarized in the table below. In 3D, the normal ordered contribution
is automatically smaller than $1/\rho_0 l^3$ since $l<\lambda$.
In 2D and 1D this is not necessarily the case; in practice 
it is convenient to adjust 
the value of $l$ so that the normal ordered contribution
is on the order of $1/\rho_0 l^d$.

What is the link with the function $g_2$ ? 
A careful calculation of $g_2$ has to involve the correction 
to the mean density due to $\hat{H}_3$ and leads to
\be
g_2(\rb) = \rho^2 + 2 \rho
\int_D \frac{d^d\kb}{(2\pi)^d} [(U_k+V_k)^2 n_k +
V_k (U_k +V_k)]
\cos(\kb\cdot\rb).
\ee
The difference between Eq.(\ref{eq:no}) and $g_2(\mathbf{0})/\rho^2-1$ therefore
involves only the small difference between the pure quasi-condensate
density $\rho_0$ and the corrected mean density.
As a consequence, the estimates at $k_B T > \mu$ also apply for
the contribution of the thermal part to $g_2(\mathbf{0})$ and give conditions
to have weak density fluctuations; in 1D, we recover the condition
obtained in the classical field model, $\chi \gg 1$, and the asymptotic
behaviour of $g_2(\mathbf{0})$, see Eq.(\ref{eq:C_large_chi}).
The quantum term leads to a divergence of $g_2(\mathbf{0})$ in
2D and 3D in the mathematical limit $l\rightarrow 0$, but this divergence
is spurious: one should not 
forget that our treatment is applicable only if $\rho_0 l^d \gg 1$.
Finally, we note that the condition to have weak phase
changes between two neighbouring cells of the grid is automatically
satisfied when $\epsilon_1 \ll 1$: one has indeed the estimate
$\epsilon_2^2 \sim 1/\rho_0 l^d$ \cite{Mora}.

\begin{center}
\begin{tabular}{|c|c|c|c|}
\hline
 & quantum term & thermal term, $k_B T<\mu$ & thermal term, $k_BT>\mu$ \\
\hline 
$d=1$  & $\frac{1}{\rho_0\xi}$ & $\frac{(k_B T/\mu)^2}{\rho_0\xi}$ &
 $\frac{k_B T/\mu}{\rho_0\xi}$\\
$d=2$  & $\frac{\ln(\xi/l)}{\rho_0\xi^2}$ & $\frac{(k_B T/\mu)^3}{\rho_0\xi^2}$
& $\frac{(k_B T/\mu)\ln(k_B T/\mu)}{\rho_0\xi^2}$ \\
$d=3$  & $\frac{1}{\rho_0\xi^2 l}$ & $\frac{(k_B T/\mu)^4}{\rho_0\xi^3}$
& $\frac{1}{\rho_0 \lambda^3}$ \\
\hline
\end{tabular}
\\
\medskip
{\small 
Estimates for  ${\langle :\delta\hat{\rho}^2(\mathbf{0}):\rangle_2}/{\rho_0^2}$.}
\end{center}

\subsubsection{Coherence properties}
The last application of the formalism
deals with the first order correlation function
of the field:
\be
g_1(\rb) \equiv \langle \hat{\psi}^\dagger(\rb) \hat{\psi}(\mathbf{0})\rangle
\simeq \langle \hat{\rho}^{1/2}(\rb)e^{i[\hat{\theta}(\mathbf{0})-
\hat{\theta}(\rb)]}\hat{\rho}^{1/2}(\mathbf{0}) \rangle.
\ee
As detailed in \cite{Mora} one first expands the two factors
$\hat{\rho}^{1/2}$ up to second order in $\delta\hat{\rho}$.
Then one calculates the thermal average with respect to the quadratic
Hamiltonian $\hat{H}_2$, proving thanks to Wick's theorem identities
like
\be
\langle e^{i \Delta \hat{\theta}}\rangle_2
= e^{-\langle(\Delta \hat{\theta})^2\rangle_2/2}
\ee
where $\Delta \hat{\theta}=\hat{\theta}(\mathbf{0})-
\hat{\theta}(\rb)$.
Then one includes to first order corrections due to $\hat{H}_3$
using Eq.(\ref{eq:tdpt}), which has the only effect of replacing
$\rho_0$ by the mean density $\rho$.
Finally one takes the continuous space limit $l\rightarrow 0$.
This leads to the result:
\be
\ln\left[g_1(\rb)/\rho\right] = \frac{-1}{\rho}
\int \frac{d^d\kb}{(2\pi)^d} \left[(U_k^2+V_k^2) n_k + V_k^2\right]
(1-\cos\kb\cdot\rb).
\ee
Remarkably this expression for the coherence function is easily related 
to the one predicted by the usual Bogoliubov theory for a Bose condensate:
\be
g_1(\rb)=\rho \exp\left[\frac{g_1^{\mathrm{Bog}}(\rb)}{\rho}-1\right].
\ee
In 3D, when a condensate is present, the argument of the exponential
is small, so that the exponential may be expanded to first order:
one recovers $g_1\simeq g_1^{\mathrm{Bog}}$.
In the infinite 1D gas, or the infinite 2D gas at finite
temperature, where no condensate is present,
the Bogoliubov prediction $g_1^{\mathrm{Bog}}(\rb)$ diverges to
$-\infty$ at large distances, whereas the prediction of
the extended Bogoliubov theory tends to zero.

The 1D case is treated in some details in \cite{Mora}.
At zero temperature, one finds that $g_1$ decays as a power law
at large distances:
\be
g_1(r) \simeq \rho \left(\frac{r_1}{r}\right)^{1/(2\pi\rho\xi)}
\label{eq:zero}
\ee
where the length scale is $r_1=e^{2-\gamma}\xi/4$, $\gamma=0.577\, 21\ldots$ 
being Euler's constant. This reproduces a result of \cite{Popov_rare}
obtained by the path integral formalism.
At finite temperature, the coherence function decays as a power law:
\be
\ln[g_1(r)/\rho] = \frac{r}{l_c} +K +o(1/r^\infty)
\label{eq:rigor}
\ee
where $o(1/r^\infty)$ is a function that tends to zero faster 
than any power law at infinity and the expression
of the temperature dependent constant $K$ is given
in \cite{Mora}. The coherence length
coincides with the prediction of
\cite{Popov};  it also coincides with the classical
field prediction (\ref{eq:k1_large_chi}), probably because the asymptotic
behaviour of $g_1$ is controlled by modes of arbitrarily low wavevectors,
and therefore of arbitrarily low Bogoliubov energy $\epsilon_k$,
for which the classicity condition $k_B T \gg \epsilon_k$ is satisfied.

An existing trap in the literature (see e.g.\cite{Schwartz})
is to calculate separately the large $r$ behaviour of the 
quantum contribution to $g_1$ (that is at $T=0$) and of the thermal
contribution to $g_1$ (that is the only involving $n_k$):
one find that the quantum bit behaves as a power law, and that
to leading order the thermal bit behaves as an exponential, so that
one may be tempted to conclude that the full $g_1$ behaves as an exponential
{\bf times} a power law, in contradiction to (\ref{eq:rigor}).
What happens in reality is that the asymptotic expansion of the
thermal bit involves a subleading power law contribution that
{\bf exactly} compensates the one of the quantum bit.

One may wonder at which distances the asymptotic behaviour
Eq.(\ref{eq:rigor}) is reached  \cite{erratum}.
The result at a temperature
$k_B T < \mu$ is that, at this critical
distance, the zero temperature asymptotics  Eq.(\ref{eq:zero})
and the finite temperature ones Eq.(\ref{eq:rigor})
should approximately give coinciding values, as justified
in \cite{these}. The critical length is therefore
\be
l_{\mathrm{crit}} \sim \lambda^2/(2\pi^2\xi)
\ee
within logarithmic accuracy, and is actually the length over
which a cross-over takes place, at a given temperature,
from a $T\simeq 0$ power law behaviour
of $g_1$ to a finite temperature $k_B T <\mu$ exponential behaviour
of $g_1$.

As a final point, we emphasize that
the present extended Bogoliubov approach gives access to the correlation
functions of the field, like $g_1$ and $g_2$, not only in the large
distance regime, but also at a length scale on the order of the
healing length $\xi$. E.g. in 1D it predicts that the third order 
derivative of $g_1$ in $r=0^+$ is non zero, irrespective of the temperature:
\be
g_1^{(3)}(0^+) = \mu^2 m^2/(2\hbar^4)
\ee
in agreement in the weakly interacting regime with an exact calculation
at zero temperature based on the Bethe ansatz \cite{BetheOlshanii}.
This is to be contrasted with other techniques like quantum hydrodynamics
or Luttinger liquids \cite{Larkin,Haldane}, 
where only the large $r$ part of the correlation
functions is obtained, with the advantage however of not being
restricted to the weakly interacting regime.

\section{Incursion in the strongly interacting regime in 1D}

The first two sections of this lecture have presented simple methods to study
weakly interacting degenerate Bose gases in arbitrary spatial dimension.
In this last section, we present simple results on the opposite regime
of extremely strong repulsive interactions, the regime of impenetrable
bosons in 1D, the so-called Tonks-Girardeau gas. 
The corresponding many-body problem
can be studied exactly by a mapping on a gas of non-interacting fermions,
a peculiarity of the 1D case.

\subsection{Mapping of impenetrable bosons onto an ideal Fermi gas}

This mapping is readily seen in the first quantization formalism for a
gas of bosons with continuous spatial coordinates and interacting
with a delta potential, $g_1 \delta(x_1-x_2)$, in the 
limit $g_1\rightarrow +\infty$ \cite{physics}. 
Such an infinitely repulsive interaction
indeed imposes that the $N$-body wavefunction vanishes when two particles
are at the same location.  On the fundamental domain of coordinates
of particles $1,\ldots,N$ sorted by ascending order, $x_1 < \ldots < x_N$,
the wavefunction of a bosonic $N$-body energy eigenstate is then an eigenstate
of the kinetic energy plus trapping potential energy operator
which vanishes at the border of the domain,
and can therefore be shown to coincide with an eigenstate of $N$ polarized
non-interacting fermions, 
which also vanishes when two particles are 
at the same location for the physically totally different
reason that it is totally antisymmetric with 
respect to any permutation of particles \cite{Girardeau1D,Lieb,Gaudin}.
Out of the fundamental domain, the bosonic wavefunction 
and the fermionic wavefunction can differ by a global sign.
As a consequence, the observables which involve the modulus squared
of the $N$-body wavefunction, like the pair correlation function
$g_2$, coincide for the bosonic and the fermionic wavefunctions;
the observables that are sensitive to the phase of the wavefunction,
like the coherence function $g_1$ or the momentum distribution, in general
do {\bf not} coincide for the bosonic and the fermionic wavefunctions.

To perform explicit calculations, a second quantized version of
the boson to fermion mapping, the so-called Jordan-Wigner transformation,
is actually convenient and easy to construct on a lattice model
\cite{Schultz,Larkin,Cazalilla}.
We therefore adapt the lattice model Eq.(\ref{eq:hamil}) to the case of 
impenetrable bosons in 1D. Since $g_0=+\infty$, configurations where two bosons
or more occupy the same lattice site are energetically suppressed, which amounts
to introducing an orthogonal projector $\cal{P}$ 
on all the configurations with at most one boson per lattice site.
At this stage, it is more convenient to manipulate the annihilation
operator $\hat{a}_x$ of one boson in the lattice site $x$ than the
field operator $\hat{\psi}(x)$, the two operators being related by
\be
\hat{\psi}(x) = \frac{1}{\sqrt{l}} \hat{a}_x
\ee
where $l$ is the grid spacing.
The infinite $g_0$ limit of the lattice Hamiltonian then reads
\be
\hat{H}_{\infty} = \sum_{x} \left[-\frac{\hbar^2}{2m}
{\cal P}\hat{a}^\dagger_x
{\cal P}\Delta\hat{a}_x{\cal P} + (U(x)-\mu)
{\cal P}\hat{a}^\dagger_x{\cal P}\hat{a}_x{\cal P}\right].
\label{eq:hinf0}
\ee
In this form, the Hamiltonian is not easy to handle as
${\cal P}\hat{a}_x{\cal P}$ and ${\cal P}\hat{a}^\dagger_x{\cal P}$
do not satisfy the usual bosonic commutation relations.
In particular one has
\be
\left({\cal P}\hat{a}_x{\cal P}\right)^2=
\left({\cal P}\hat{a}_x^\dagger{\cal P}\right)^2=0 
\ee
\be
\{{\cal P}\hat{a}_x{\cal P},{\cal P}\hat{a}^\dagger_x{\cal P}\}=1
\ee
which is reminiscent of a fermionic system.
However it is not quite a fermionic system because the 
${\cal P}\hat{a}{\cal P}$'s 
at different lattice sites commute rather than anticommute.
One then introduces
\be
\hat{c}_x \equiv 
\exp\left[i\pi\sum_{y<x} \hat{a}_y^\dagger\hat{a}_y\right]
{\cal P} \hat{a}_x {\cal P}
\ee
or equivalently
\be
{\cal P} \hat{a}_x {\cal P} = 
\exp\left[i\pi\sum_{y<x} \hat{c}_y^\dagger\hat{c}_y\right]
\hat{c}_x
\ee
and one checks that the $\hat{c}_x$ and $\hat{c}_x^\dagger$ satisfy
the usual fermionic anticommutation relations:
\bea
\{\hat{c}_x,\hat{c}_y\} &=& 0 \\
\{\hat{c}_x,\hat{c}_y^\dagger\} &=& \delta_{x,y}.
\eea
After this transform, and if one uses the simplified representation
Eq.(\ref{eq:3pt}) of the discrete Laplacian, the model Hamiltonian
Eq.(\ref{eq:hinf0}) becomes
\be
\hat{H}_{\infty} = \sum_x
-\frac{\hbar^2}{2m} \hat{c}_x^\dagger \Delta \hat{c}_x
+(U(x)-\mu) \hat{c}_x^\dagger \hat{c}_x
\label{eq:hinf}
\ee
which is a Hubbard model for non-interacting fermions.
We used the fact that 
\be
{\cal P} \hat{a}^\dagger_x {\cal P} \hat{a}_{x-l} {\cal P}
= \hat{c}^\dagger_x e^{-i\pi \hat{c}^\dagger_{x-l} \hat{c}_{x-l}}
\hat{c}_{x-l} =\hat{c}^\dagger_x \hat{c}_{x-l}.
\ee
One is left with a quadratic fermionic Hamiltonian which can be
diagonalized \cite{difficile}.
In the subsections to come, we deduce some observables of the
impenetrable Bose gas.

\subsection{Pair correlation function of impenetrable bosons}

The second order correlation function of the bosonic field, defined
in Eq.(\ref{eq:def_g2}), is found to coincide with the one of the
non-interacting fermions, since ${\cal P} \hat{a}^\dagger_x\hat{a}_x
{\cal P} = \hat{c}^\dagger_x\hat{c}_x$:
\be
g_2(x) = \frac{1}{l^2} \langle \hat{c}^\dagger_x \hat{c}^\dagger_0
\hat{c}_0 \hat{c}_x \rangle.
\ee
The gas is at thermal equilibrium with temperature $T$. Wick's theorem
can be used for the fermions to express $g_2$ in terms of the first order
coherence function $f(x)$ of the fermions, and then in terms of
the occupation
numbers of the fermionic eigenmodes. For a spatially homogeneous system
of density $\rho$ and
in the thermodynamic limit $\rho L\rightarrow +\infty$
and continuous limit $\rho l\rightarrow 0$ one finds
\be
g_2(x) = \rho^2 - |f(x)|^2
\ee
with
\be
f(x) = \int_{-\infty}^{+\infty} \frac{dk}{2\pi} \frac{e^{-ikx}}
{\exp[\beta(\hbar^2k^2/(2m)-\mu)]+1}.
\label{eq:deff}
\ee
As expected for impenetrable bosons or for polarized fermions, $g_2(0)=0$.
At zero temperature, 
\be
f(x)= \frac{\sin k_F x}{\pi x}
\ee
where $k_F= \pi\rho$ is the Fermi wavevector, so that the size of the
hole of $g_2$ is the mean interparticle separation $1/\rho$.
In the non-degenerate regime $T>T_F$ the hole size scales as the
thermal de Broglie wavelength.

An interesting application of $g_2$ is to calculate the fluctuations of
the number $N_X$ of impenetrable bosons in some sub-interval of length
$X$ of the bulk gas. The expectation value of $N_X$ is $\rho X$
and its variance is
\be
\mbox{Var}N_X = \rho X+ \int_0^X dx\, \int_0^X dx'\, [g_2(x-x')-\rho^2]
.
\ee
A more tractable formula involving only a single spatial integral is
derived as follows:
\be
\int_0^X dx\, \int_0^X dx' |f(x'-x)|^2  = \int_0^X dx\, 1\times
\int_{-x}^{X-x} dx' |f(x')|^2 
\ee
and then one integrates by part in the integral over $x$, taking
the derivative of $\int_{-x}^{X-x} dx'$ and integrating the factor $1$.
This leads to
\be
\mbox{Var}N_X = \rho X -2 \int_0^X dx\, (X-x) |f(x)|^2.
\label{eq:vargen}
\ee

At $T=0$, an explicit expression is then obtained using the special functions
$\mathrm{Si}(x)$ and $\mathrm{Ci}(x)$:
\be
\mbox{Var}N_X = -\frac{1}{\pi^2}\left[
-\pi k_F X +\cos(2k_F X)-1 +2 k_F X\, \mathrm{Si}(2 k_F X)-\gamma
-\ln(2k_F X)+ \mathrm{Ci}(2k_F X)
\right]
\ee
from which the large $k_F X$ behaviour follows:
\be
\mbox{Var}N_X = \frac{\gamma+1+\ln 2k_F X}{\pi^2}+O(1/(k_F X)^2)
\label{eq:var0}
\ee
where $\gamma=0.57721\ldots$ is Euler's constant.
This results in extremely weak relative fluctuations of the number
of particles, well below the Poisson limit for $k_F X \gg 1$, since the
dependence with $X$ is logarithmic.

At finite temperature, the Fermi distribution is a 
smooth, $C^\infty$ function of the momentum and decreases faster than an
exponential at infinity. One can then show, by repeated
integration by parts in Eq.(\ref{eq:deff}), 
integrating $\exp(-ikx)$ with respect to $k$,
that $f(x)$ decreases faster than any power law at $x=\infty$.
As a consequence, in the limit of a large interval,
\be
\mbox{Var}N_X = \left[\rho-\int_{-\infty}^{+\infty} dx 
\, |f(x)|^2\right]X + 
\int_{-\infty}^{+\infty}dx\, |x|\, |f(x)|^2 + o(1/X^\infty)
\ee
where the remainder $o(1/X^\infty)$ decreases faster than any power law.
From the Parseval-Plancherel identity one finds the more illuminating form
\be
\mbox{Var}N_X \sim X \int_{-\infty}^{+\infty} \frac{dk}{2\pi} n(k)\left[
1-n(k)\right] 
\label{eq:varT}
\ee
where $n(k)$ is the Fermi distribution function of the ideal Fermi gas.
This coincides with the usual grand canonical result for a
system of total length $X$ treated in the thermodynamic limit.
In the non-degenerate regime, $n(k)$ can be neglected as compared to
$1$ and the variance of $N_X$ is the one of Poisson statistics, as expected
for independent classical particles.

\begin{figure}[htb]
\centerline{\includegraphics[height=6cm,clip]{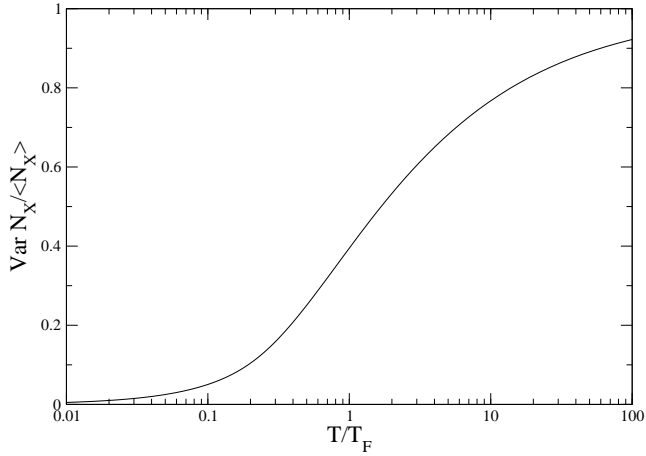}}
\caption{In the limit of a large interval of length $X$, ratio of
the variance and the mean value of the number of impenetrable
bosons in this interval, or equivalently of the number of fermions for
an ideal Fermi gas, as function of the temperature in units
of the Fermi temperature.}
\label{fig:var}
\end{figure}

What happens in the opposite regime $T\ll T_F$~?
From the identity
\be
\partial_{\mu} n(k) = \beta n(k) \left[1-n(k)\right]
\ee
and the low temperature expansion of the density of a 1D ideal Fermi gas
at fixed chemical potential \cite{Diu}:
\be
\rho = \frac{1}{\pi} \left(\frac{2m\mu}{\hbar^2}\right)^{1/2}
\left[1  - \frac{\pi^2}{24}\left(\frac{k_B T}{\mu}\right)^2
+O(T^4)
\right]
\ee
one obtains the low temperature expansion
\be
\mbox{Var}N_X \sim \rho X \left[\frac{T}{2T_F}+\frac{\pi^2}{24}
\left(\frac{T}{T_F}\right)^3+O((T/T_F)^5)\right]
\ee
where $k_B T_F = \hbar^2 k_F^2/2m$ is the Fermi energy \cite{concavity}.
The fluctuations of the number of impenetrable bosons in a large
interval is therefore subpoissonian at $0<T\ll T_F$, but still proportional
to $X$.
We have plotted in figure \ref{fig:var} the ratio of the variance
of $N_X$ and of its mean value, in the large $X$ limit, as function
of $T/T_F$.

For a fixed temperature $T\ll T_F$, one can find the critical length
$X_c$ below which the variance of the number of particles 
has the zero temperature behaviour Eq.(\ref{eq:var0})
and above which it acquires the finite temperature behaviour 
Eq.(\ref{eq:varT}). A first way of finding $X_c$ is to use
Eq.(20) of \cite{Larkin} giving an approximation to the function
$f(x)$ at $T\ll T_F$:
\be
|f(x)|^2 \simeq k_F^2
\frac{T^2}{4 T_F^2} \frac{\sin^2 k_Fx}{\sinh^2(\pi T k_F x/2 T_F)}.
\ee
One sees that $f(x)$ differs weakly from its zero temperature value
as long as $\pi T k_F |x| \ll 2 T_F$. One the contrary, if
$\pi T k_F |x| \gg 2 T_F$, $|f(x)|^2$ is exponentially small
so that one is left with Eq.(\ref{eq:varT}).  This gives a cross-over length
\be
k_F X_c \sim \frac{2T_F}{\pi T}.
\ee
A second way of finding $X_c$ is to equate Eq.(\ref{eq:var0})
and Eq.(\ref{eq:varT}). Assuming $\ln 2k_F X_c \sim 1$, this gives
the same order of magnitude for $X_c$ as in the previous equation.
A third way is to imagine that one can introduce a fictitious box
of size $X$ with periodic boundary conditions, containing a fictitious
Fermi gas with the same chemical potential as the bulk system. 
This box gives wrong predictions at zero temperature, as the variance of
$N_X$ would then be exactly zero, but gives indeed the correct result
Eq.(\ref{eq:varT}) for finite $T$,
when the thermodynamic limit approximation
is applicable for the fictitious box, which occurs when
\be
k_B T > \Delta E
\ee
where $\Delta E$ is the energy separation between two consecutive energy
levels in the box around the Fermi energy. One finds $\Delta E \sim
2\pi \hbar^2 k_F/mX$ which leads to the same $X_c$ as before, within
a numerical factor.

\subsection{First order coherence function of impenetrable bosons}
The first order coherence function in the general case of a spatially
inhomogeneous system is defined as
\be
g_1(x,y)=\langle \hat{\psi}^\dagger(y)\hat{\psi}(x)\rangle=\frac{1}{l}\langle \hat{a}^\dagger_y\hat{a}_x\rangle.
\ee
As the expectation value is taken over a density operator $\sigma$ 
such that ${\cal P}\sigma{\cal P}=\sigma$, one can add factors equal
to ${\cal P}$ in the above expression. This leads to an expression
for $g_1$ in terms of the fermionic operators:
\be
g_1(x,y)=-l^{-1}\langle\hat{c}_y^\dagger\hat{c}_x 
e^{i\pi\sum_{z=x}^{y}\hat{c}_z^\dagger\hat{c}_z}\rangle 
\label{eq:g1f}
\ee
where we have assumed that $x < y$ \cite{help}.

We now go through a sequence of transformations of Eq.(\ref{eq:g1f}).
First we rewrite it as
\be
g_1(x,y)= -l^{-1}
\frac{\mathrm{Tr}\left[\hat{c}_y^\dagger\hat{c}_x e^A e^B\right]}
{\mathrm{Tr}\left[e^B\right]}
\ee
where $A$ and $B$ are operators that are quadratic in the fermionic variables.
$B$ is equal to $-\beta \hat{H}_{\infty}$, where $\beta=1/k_B T$ and
the Hamiltonian $\hat{H}_\infty$ is given by Eq.(\ref{eq:hinf}). The operator $A$
is $i\pi$ times the operator counting the number of fermions on the
sites from $x$ to $y$. We introduce the matrices in the lattice basis
of the one-body operators corresponding to $A$ and $B$:
\bea
\label{eq:A}
A &=& \sum_{\alpha,\beta} {\cal A}_{\alpha\beta} \hat{c}^\dagger_\alpha
\hat{c}_\beta \\
\label{eq:B}
B &=& \sum_{\alpha,\beta} {\cal B}_{\alpha\beta} \hat{c}^\dagger_\alpha
\hat{c}_\beta 
\eea
Let us introduce a matrix ${\cal C}$ such that 
\be
e^{\cal A} e^{\cal B} = e^{\cal C}.
\label{eq:matabc}
\ee
Then, as we show in the appendix \ref{appen:ABC},
$e^A e^B = e^C$
where the operator $C$ is defined as
\be
C= \sum_{\alpha,\beta} {\cal C}_{\alpha\beta} \hat{c}^\dagger_\alpha
\hat{c}_\beta.
\label{eq:C}
\ee
We also derive in this appendix the following expressions:
\be
\frac{\mathrm{Tr}\left[\hat{c}_y^\dagger\hat{c}_x e^C\right]}
{\mathrm{Tr}\left[e^C\right]}=
\left(\frac{1}{1+e^{-{\cal C}}}\right)_{xy}.
\label{eq:onebodyc}
\ee
\be
\mathrm{Tr}\left[e^C\right]=\mathrm{det}\,(1+e^{\cal C}).
\ee
What remains to be done is to eliminate $\cal C$ in terms of $\cal A$
and $\cal B$: this is easy since $\cal C$ appears in the final result
only through its exponential, so that Eq.(\ref{eq:matabc}) can be used
directly. A further simplification arises from the fact that
$\cal A$ is proportional to a projector. The calculations are detailed
in the appendix, we give the result:
\be
g_1(x,y) = \frac{1}{2l}\left(\frac{1}{1-2{\cal S}}\right)_{xy}
\mathrm{det}(1-2{\cal S}) \ \ \ \ \ \ \ \mathrm{for} \ \ \ x<y
\label{eq:g1kor}
\ee
where the matrix ${\cal S}$ is defined on the lattice sites in between $x$
and $y$ by the following thermal averages:
\be
S_{\alpha\beta}\equiv\langle\hat{c}^\dagger_\beta\hat{c}_\alpha\rangle
\ \ \ \ \ \mathrm{for} \ \ x\leq \alpha,\beta\leq y.
\ee
Interestingly, for the spatially homogeneous case,
in the continuous limit $l\rightarrow 0$, 
where matrices are replaced by operators acting on a functional space and
the dispersion relation for fermions is quadratic,
Eq.(\ref{eq:g1kor}) becomes formally equivalent to the expression
of $g_1$ given in \cite {Korepine}
for impenetrable bosons in the Lieb-Liniger model.
In the general case, several equivalent forms of (\ref{eq:g1kor}) can be obtained after simple linear algebra
manipulations.
First, using the matrix identity $M^{-1}= {}^t \mathrm{com}\, M/\mathrm{det}
\, M$ relating the inverse of a matrix $M$ to the transpose of its comatrix,
we obtain:
\be
g_1(x,y) = \frac{1}{2l} \mathrm{det} 
(2{\cal S}_{\alpha\beta}-\delta_{\alpha,\beta})|_{x\leq \alpha < y,x<\beta\leq y} \ \ \ \ \ \mathrm{for} \ \ x<y
\label{eq:g1useful}
\ee
that is in terms of the determinant of the matrix obtained
by suppressing the first column and the last line of $2{\cal S}-1$.
This coincides with \cite{Schultz} and,
within a factor of two, with Eq.(8) of \cite{Larkin}.
Second, introducing the one-fermion density operator $\hat{\sigma}$ such that $\langle z|\hat{\sigma}|z'\rangle=
\langle \hat{c}^\dagger_{z'} \hat{c}_{z}\rangle/l$ due to the normalisation (\ref{eq:norma}),
and the single-particle orthogonal projector $\hat{\Pi}$ over the discrete position interval $x\leq z\leq y$,
one gets the operatorial form $g_1(x,y)=\frac{1}{2} \langle x|(1-2\hat{\Pi}\hat{\sigma}\hat{\Pi})^{-1}|y\rangle
\, \mathrm{det}(1-2\hat{\Pi}\hat{\sigma}\hat{\Pi})$ that can be rewritten as
\be
g_1(x,y)=\langle x|\hat{\sigma}^{1/2}(1-2\hat{R})^{-1}\hat{\sigma}^{1/2}|y\rangle \,\mathrm{det}(1-2\hat{R})
\ \ \ \ \ \mathrm{for} \ \ x<y
\label{eq:g1R}
\ee
with $\hat{R}=\hat{\sigma}^{1/2}\hat{\Pi}\,\hat{\sigma}^{1/2}$. One has taken advantage of the identity
\be
(1-2\hat{\Pi}\hat{\sigma}\hat{\Pi})^{-1}-1=2\hat{\Pi}\, \hat{\sigma}^{1/2} (1-2\hat{R})^{-1} \hat{\sigma}^{1/2} \hat{\Pi}
\ee
that results from a geometric series expansion and from $\hat{\Pi}^2=\hat{\Pi}$, 
$\hat{\sigma}^{1/2} (\hat{\Pi}\hat{\sigma}\hat{\Pi})^n\hat{\sigma}^{1/2}=\hat{R}^{n+1}$
$\forall n>0$. Similarly $\mathrm{Tr}[(\hat{\Pi}\hat{\sigma}\hat{\Pi})^n]
=\mathrm{Tr}(\hat{R}^n)$, so that $\mathrm{det}(1-2\hat{\Pi}\hat{\sigma}\hat{\Pi}) =\mathrm{det}(1-2\hat{R})$
due the series expansion $\ln\mathrm{det}(1-\lambda M)=-\sum_{n\geq 1} 
\frac{\lambda^n}{n} \mathrm{Tr}(M^n)$ valid for any matrix $M$. 
If one writes (\ref{eq:g1R}) in the eigenbasis of $\hat{\sigma}$, one obtains a finite temperature generalisation 
of \cite{Pezer}; this was pointed out to us by Yasar Atas and Isabelle Bouchoule, who obtained an independent
derivation.

Extracting analytically physical information for $g_1$ is quite technical.
The key result is the absence of long range order for the homogeneous system
at zero temperature \cite{Lenard,Tracy}:
\be
g_1(X)\sim \frac{A\rho}{|k_F X|^{1/2}}
\label{eq:asympt}
\ee
where $A=0.92418\ldots$ is a numerical constant and $k_F X \gg 1$.
We refer to \cite{Gangardt} for an overview of the various analytical
results.
An important consequence of Eq.(\ref{eq:asympt}) is that the momentum distribution
of impenetrable bosons at $T=0$, which is the Fourier transform of $g_1$,  
$\pi(k)=\int_{-\infty}^{+\infty} \cos(kx) g_1(x)$, diverges as 
$2A[k_F/(2\pi k)]^{1/2}$ when $k\rightarrow 0$.

\begin{figure}[htb]
\centerline{\includegraphics[height=6cm,clip]{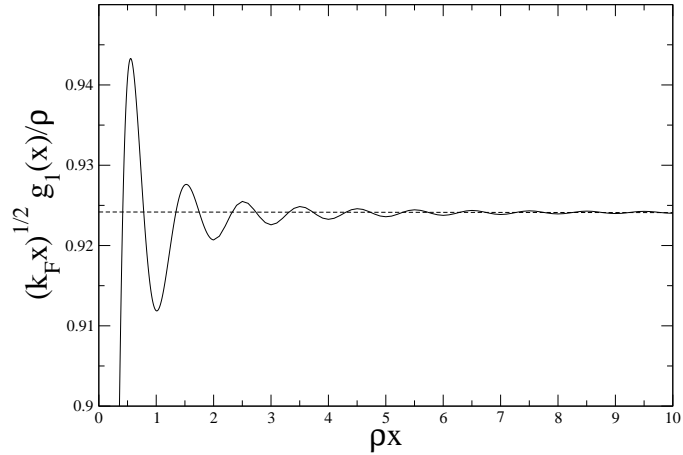}}
\caption{Coherence function $g_1(x)$ of a zero temperature
spatially 
homogeneous gas of 1D impenetrable bosons in the thermodynamical limit,
calculated numerically from Eq.(\ref{eq:g1useful}),
with $\rho l=0.005$. $g_1$ was multiplied
by $(k_F x)^{1/2}$, where the Fermi wavevector $k_F$ is
related to the density $\rho$ by $k_F = \pi \rho$, to reveal
that $g_1$ oscillates around its asymptotic value.
Dashed line: constant $A$ of Eq.(\ref{eq:asympt}).
}
\label{fig:g1b}
\end{figure}

In figure \ref{fig:g1b} is plotted the result of a calculation
of $g_1(X)$ based on a numerical evaluation of the determinant
in Eq.(\ref{eq:g1useful}) at $T=0$. We have actually plotted the
product of $g_1(X)$ with $(k_FX)^{1/2}$ to reveal that 
$g_1$ exhibits damped oscillations around the asymptotic formula
Eq.(\ref{eq:asympt}): on a direct plot of the bare $g_1$, these
oscillations are not visible in practice. 
Note that these oscillations
have a maximum roughly at the half integer values of $\rho X$ and have a minimum
roughly at the integer values of $\rho X$: this is in contradiction with the
asymptotic series expansion of \cite{Tracy} but is reproduced
by the analytical approach of \cite{Gangardt}.

Can we understand the emergence of such a $1/X^{1/2}$ asymptotic behaviour of $g_1$
at zero temperature~?
We present here an interpretation with no pretension of rigor. 
The first step is to realize that the power law of the long range behavior of $g_1$ 
is not changed if one eliminates the factor 
$\hat{c}_y^\dagger\hat{c}_x/l$ in the Eq.(\ref{eq:g1f}). This is suggested by the
calculations of \cite{Schultz}, and we have checked this fact numerically.
What remains to be understood is the long range behaviour of the function
\be
G(X) \equiv \langle 
e^{i\pi \hat{N}_X} \rangle
= \sum_{n=0}^{+\infty} \pi_n e^{i\pi n}
\ee
where $\hat{N}_X=\sum_{z=0}^{X} \hat{c}^\dagger_z \hat{c}_z$ is the operator
giving the number of particles in the considered interval of length $X$, and $\pi_n$
is the probability of having $n$ particles  in the interval.
$G(X)$ is then the difference of the probability of having an even number of particles
and the probability of having an odd number of particles in the interval.
This can be estimated heuristically by assuming that the probability distribution 
of the number of fermions in the interval is roughly Gaussian for $k_F X \gg 1$
\cite{Lesovik}.
One can then extend the summation for $n$ to $-\infty$ since the standard deviation $\sigma_X$
of the Gaussian, while being much larger than unity, is much smaller
than the mean of the Gaussian.
Using the Poisson formula
\be
\sum_{n=-\infty}^{+\infty} f(n) = \sum_{n=-\infty}^{+\infty} \tilde{f}(2\pi n)
\ee
where $\tilde{f}(k)=\int dx f(x) \exp(-ikx)$ is the Fourier transform of an arbitrary
function $f(x)$, and restricting in the right hand side sum to the leading terms
with $n=0$ and $n=1$, one obtains
\be
G(X) \simeq 2\cos(\pi\rho X) e^{-\pi^2\sigma_X^2/2}.
\ee
The expression Eq.(\ref{eq:var0}) of the standard deviation in the large $X$ limit leads to
\be
G(X) \simeq \frac{0.36}{(\rho X)^{1/2}}\cos(\pi\rho X).
\ee
A numerical calculation confirms this result, with a coefficient $0.33$ rather than $0.36$.

The conclusion is that the $1/X^{1/2}$ behaviour of the first order coherence
function in the ground state of impenetrable bosons in 1D reflects a property
of the counting statistics of a 1D zero temperature ideal Fermi gas in an interval of length $X$.

\Appendix
\section{Wick's theorem}\label{appen}

Consider the following problem: for a density operator (here for bosons
for simplicity, we shall come to the case of fermions later)
\be
\rho=\frac{1}{Z} \exp\left[-\sum_\alpha \nu_\alpha \hat{a}_\alpha^\dagger
\hat{a}_\alpha \right]
\label{eq:dens}
\ee
where all $\nu_\alpha$ are strictly positive, and the $\hat{a}_\alpha$'s obey
bosonic commutation relations:
\be
[\hat{a}_\alpha,\hat{a}_\beta^\dagger] = \delta_{\alpha\beta}
\ee
\be
[\hat{a}_\alpha,\hat{a}_\beta] = 0,
\ee
calculate the expectation value
\be
I = \langle \hat{b}_1\ldots\hat{b}_{2n}\rangle 
\equiv \mbox{Tr}\left[ \hat{b}_1\ldots\hat{b}_{2n}
\frac{1}{Z} e^{-\sum_\alpha \nu_\alpha \hat{a}_\alpha^\dagger
\hat{a}_\alpha} \right]
\ee
where the $\hat{b}_i$'s are arbitrary linear combinations of $\hat{a}_\alpha$'s
and $\hat{a}_\alpha^\dagger$'s.
We consider an even number of factors as the expectation value of the
product of an odd number of $\hat{b}$'s vanishes.

A physical example is simply the ideal Bose gas at thermal equilibrium in the 
grand canonical ensemble, with $\nu_\alpha= \beta(\epsilon_\alpha-\mu)$.
Note that the density operator in the canonical ensemble does not reduce
to (\ref{eq:dens}) as it involves in addition the projector $P_N$ on the 
subspace with a fixed total number of particles equal to $N$.

The explicit calculation proceeds in four steps, in a derivation inspired 
by the lectures on statistical physical of Jacques des Cloizeaux
at the Ecole normale sup\'erieure of Paris, France, in 1988.

{\bf First step}: assume that $\hat{b}_1$ is either $\hat{a}_{\alpha_0}$
or $\hat{a}_{\alpha_0}^\dagger$. Then there exists a number $\lambda$
such that
\be
\label{eq:through}
\hat{\rho} \hat{b}_1 = \lambda \hat{b}_1 \hat{\rho}.
\ee
One can check this identity using the Fock basis. E.g. one finds
$\lambda=\exp\nu_{\alpha_0}$ for $\hat{b}_1=\hat{a}_{\alpha_0}$.
By using the invariance of the trace by a cyclic permutation of the operators,
one transfers $\hat{b}_1$ to the extreme right and one then puts it through
$\hat{\rho}$ thanks to (\ref{eq:through}):
\be
\langle \hat{b}_1\ldots\hat{b}_{2n}\rangle =
\lambda \langle \hat{b}_2\ldots\hat{b}_{2n}\hat{b}_1\rangle.
\ee
The following commutation relation allows to reintegrate the factor
$\hat{b}_1$ in the extreme left:
\be
[\hat{b}_1, \hat{b}_2\ldots\hat{b}_{2n}] =
\sum_{j=2}^{2n} 
\hat{b}_{1}\ldots \hat{b}_{j-1}
[\hat{b}_1,\hat{b}_j] 
\hat{b}_{j+1}\ldots \hat{b}_{2n}.
\ee
Since the commutator of two $\hat{b}$'s is a pure number, we obtain
\be
\langle \hat{b}_1\ldots\hat{b}_{2n}\rangle =
\sum_{j=2}^{2n} \frac{[\hat{b}_1,\hat{b}_j]}{\lambda-1}
\langle \hat{b}_2\ldots\hat{b}_{j-1} \hat{b}_{j+1}\ldots \hat{b}_{2n}\rangle.
\ee

{\bf Second step}: Apply this general formula for two operators:
\be
\langle\hat{b}_1\hat{b}_j \rangle
= \frac{[\hat{b}_1,\hat{b}_j]}{\lambda-1}.
\ee
This identity, which in particular implies the Bose formula,
allows to rewrite the general formula in a simpler way, without any
apparition of $\lambda$:
\be
\langle \hat{b}_1\ldots\hat{b}_{2n}\rangle =
\sum_{j=2}^{2n} \langle\hat{b}_1\hat{b}_j \rangle
\langle \hat{b}_2\ldots\hat{b}_{j-1} \hat{b}_{j+1}\ldots \hat{b}_{2n}\rangle.
\label{eq:recur}
\ee

{\bf Third step}: the linearity of (\ref{eq:recur}) with respect to $\hat{b}_1$
implies that it holds even if $\hat{b}_1$ is an arbitrary linear combination
of $\hat{a}_\alpha$'s and $\hat{a}_\alpha^\dagger$'s!

{\bf Fourth step}: iterate the formula (\ref{eq:recur}) down to $2n=2$.
So the most general expectation value $I$ can be expressed in terms of expectation values
of binary products, which are of course known from the Bose formula.

The result is found to have a simple structure if one introduces the concept of a
contraction. One collects the first factor $\hat{b}_1$ with one of the other factors,
that we call $\hat{b}_\alpha$: one says that one performs a {\it contraction}
of $\hat{b}_1$ with $\hat{b}_\alpha$. One contracts the next factor available,
$\hat{b}_\beta$, with one of the factors left, $\hat{b}_\gamma$. Note that
$\hat{b}_\beta=\hat{b}_2$ if $\alpha\ne 2$, otherwise $\beta=3$. One repeats this
process until all the factors are contracted. Then
\be
\langle \hat{b}_1\ldots\hat{b}_{2n}\rangle = \sum_{\rm all\ contractions}
\langle \hat{b}_1\hat{b}_\alpha\rangle \langle\hat{b}_\beta\hat{b}_\gamma\rangle
\ldots \langle\hat{b}_\psi\hat{b}_\omega\rangle.
\ee

Let us count the number of possible ways of contracting the $2n$ factors.
There are $2n-1$ possibilities for the choice of the companion of $\hat{b}_1$,
that is for the choice of $\hat{b}_\alpha$. There are $2n-3$ possibilities for the choice of the 
companion of $\hat{b}_\beta$, that is for the choice of $\hat{b}_\gamma$... Finally,
there is a single possibility for the choice of the companion of $\hat{b}_\psi$, 
that is for the choice of $\hat{b}_\omega$. The total number of possible contractions is
therefore
\be
(2n-1)\times(2n-3)\times\ldots\times 1=\frac{(2n)!}{2^n n!}.
\ee

Let us give an example for $n=2$:
\be
\langle\hat{b}_1\hat{b}_2\hat{b}_3\hat{b}_4\rangle =
\langle \hat{b}_1\hat{b}_2\rangle \langle \hat{b}_3\hat{b}_4\rangle
+\langle \hat{b}_1\hat{b}_3\rangle \langle \hat{b}_2\hat{b}_4\rangle
+\langle \hat{b}_1\hat{b}_4\rangle \langle \hat{b}_2\hat{b}_3\rangle.
\ee

What happens for fermions? The annihilation and creation operators
now obey anticommutation relations
\bea
\{\hat{a}_\alpha,\hat{a}_\beta^\dagger\} &=& \delta_{\alpha\beta} \\
\{\hat{a}_\alpha,\hat{a}_\beta\} &=& 0.
\eea
As the number operator $\hat{a}_\alpha^\dagger\hat{a}_\alpha$
is now bounded from above by unity, the coefficient $\nu_\alpha$
is no longer restricted to positive values.
The same technique as for bosons can be applied to derive the following recursive
relation:
\be
\langle \hat{b}_1\ldots\hat{b}_{2n}\rangle =
\sum_{j=2}^{2n} (-1)^j\langle\hat{b}_1\hat{b}_j \rangle
\langle \hat{b}_2\ldots\hat{b}_{j-1} \hat{b}_{j+1}\ldots \hat{b}_{2n}\rangle.
\ee
The factor $(-1)^j$ appears because $\hat{b}_j$ `crossed' $j-2$ factors.
This gives rise to the Wick's rule:
\be
\langle \hat{b}_1\ldots\hat{b}_{2n}\rangle = \sum_{\rm all\, contractions}
\epsilon_{\rm contraction}
\langle \hat{b}_1\hat{b}_\alpha\rangle \langle\hat{b}_\beta\hat{b}_\gamma\rangle
\ldots \langle\hat{b}_\psi\hat{b}_\omega\rangle.
\ee
A sign $\pm$, that we denote as $\epsilon_{\rm contraction}$, is associated to
each contraction. Each contraction actually defines a permutation of the $2n$
factors, performing the following mapping:
\be
(1,2,3,4,\ldots,2n-1,2n) \rightarrow (1,\alpha,\beta,\gamma,\ldots,\psi,\omega).
\ee
$\epsilon_{\rm contraction} $ is simply the signature of this permutation, that is
$(-1)$ to the power the number of transpositions of two elements required
to realize the permutation.
Let us give an example for $n=2$:
\be
\langle\hat{b}_1\hat{b}_2\hat{b}_3\hat{b}_4\rangle =
\langle \hat{b}_1\hat{b}_2\rangle \langle \hat{b}_3\hat{b}_4\rangle
-\langle \hat{b}_1\hat{b}_3\rangle \langle \hat{b}_2\hat{b}_4\rangle
+\langle \hat{b}_1\hat{b}_4\rangle \langle \hat{b}_2\hat{b}_3\rangle.
\ee

\section{Some useful identities for Gaussian operators}
\label{appen:ABC}

Consider operators $A$, $B$ and $C$ that are quadratic in the fermionic
field, in the sense defined by Eq.(\ref{eq:A},\ref{eq:B},\ref{eq:C}).
The square matrices $\cal A$, $\cal B$ and $\cal C$ of the associated quadratic
forms are linked by Eq.(\ref{eq:matabc}).

\subsection{$ABC$ identity}
We prove that the identity $e^A e^B= e^C$ holds.
The idea of the proof is to consider each exponential factor as an
evolution operator and to show that the composition of two Gaussian
evolution operators gives a Gaussian evolution operator.

Let us introduce the formal time evolution:
\be
\hat{c}_\alpha(\tau) \equiv e^{\tau B} \hat{c}_\alpha e^{-\tau B}.
\ee
The time derivative of $\hat{c}_\alpha(\tau)$ involves the commutator of $B$
and $\hat{c}_\alpha$ that can be calculated from the fermionic
anticommutation relations:
\be
\frac{d}{d\tau} \hat{c}_\alpha(\tau) = -\sum_\beta {\cal B}_{\alpha\beta}
\hat{c}_\beta(\tau).
\ee
This differential system with the initial conditions
$\hat{c}_\alpha(0)=\hat{c}_\alpha$ is integrated formally 
in terms of the evolution matrix $e^{-{\cal B}\tau}$. Specializing
to $\tau=1$ we obtain
\be
e^{B} \hat{c}_\alpha e^{-B} = \sum_\beta 
\left(e^{-{\cal B}}\right)_{\alpha\beta} \hat{c}_\beta.
\label{eq:useful}
\ee

By using the same type of identity with $B$ replaced with $A$ or
with $C$ we obtain
\bea
e^{A} e^{B} \hat{c}_\alpha e^{-B} e^{-A}  &=& 
\sum_{\beta,\gamma} \left(e^{-{\cal B}}\right)_{\alpha\beta}
\left(e^{-{\cal A}}\right)_{\beta\gamma} \hat{c}_\gamma\\
&=& \sum_{\gamma} \left(e^{-{\cal B}}e^{-{\cal A}}\right)_{\alpha\gamma}
\hat{c}_\gamma
= \sum_{\gamma} \left( e^{-{\cal C}}\right)_{\alpha\gamma} \hat{c}_\gamma \\
&=&  e^C \hat{c}_\alpha e^{-C}
\eea
where we used Eq.(\ref{eq:matabc}).
Similarly one can show that
\be
e^{A} e^{B} \hat{c}_\alpha^\dagger e^{-B} e^{-A} 
= e^C \hat{c}_\alpha^\dagger e^{-C}.
\ee

Consider now the action of $e^C$ on an Hartree-Fock state defined by an
arbitrary list $\gamma_1,\ldots,\gamma_n$ of single particle states:
\be
|\psi\rangle \equiv \hat{c}_{\gamma_1}^\dagger \ldots 
\hat{c}_{\gamma_n}^\dagger |0\rangle.
\ee
We perform the following sequence of transformations:
\bea
e^C |\psi\rangle &=& \left(e^C \hat{c}_{\gamma_1}^\dagger e^{-C}\right)\ldots 
\left(e^C\hat{c}_{\gamma_n}^\dagger e^{-C}\right) |0\rangle \\
&=& 
\left(e^A e^B \hat{c}_{\gamma_1}^\dagger e^{-B} e^{-A}\right)\ldots
\left(e^A e^B \hat{c}_{\gamma_n}^\dagger e^{-B} e^{-A} \right) |0\rangle \\
&=& e^A e^B |\psi\rangle
\eea
where we used the fact that $e^{-C}|0\rangle=e^{-B} |0\rangle= e^{-A} |0\rangle
=|0\rangle$ as can
be checked by the power series expansion of the exponential.
The operators $e^C$ and $e^A e^B$ have the same action on all the
Hartree-Fock states, which form a basis, so they are equal.

\subsection{Trace and one-body expectation values of a Gaussian operator}
What is the trace of $e^C$~? What is the one-body operator
corresponding to $e^C$~? 
We give the answer assuming that the
matrix $\cal C$ is diagonalizable, which is sufficient to conclude in the 
general case using a continuity argument.

We call $e_i$ the set of eigenvectors of $\cal C$, with eigenvalue
${\cal C}_i$, so that ${\cal C} e_i = {\cal C}_i e_i$, and we call $f_j$ the adjoint basis
normalized such that
\be
f_j^* \cdot e_i = \delta_{i,j}
\label{eq:adj}
\ee
so that the closure relation is satisfied:
\be
\sum_i (e_i)_\alpha (f_i)^*_\beta = \delta_{\alpha,\beta}
\label{eq:clos}
\ee
where $(e_i)_\alpha$ is the component of the vector $e_i$ in the
original single particle basis.

We introduce the corresponding creation and annihilation operators:
\bea
\hat{b}_i &\equiv& \sum_\alpha \hat{c}_\alpha (f_i)_\alpha^* \\
\hat{b}_i^+ &\equiv& \sum_\alpha \hat{c}_\alpha^\dagger (e_i)_\alpha.
\eea
Note that in general the creation operator $\hat{b}_i^+$ is {\bf not}
the Hermitian conjugate of the annihilation operator $\hat{b}_i$
since the matrix $\cal C$ is not necessarily Hermitian.
This is why we introduced the symbol $+$ rather than $\dagger$ in the
exponent.
However this has little effect on the algebra of second quantization!
The $\hat{b}_i$'s obviously anticommute among themselves, since they
are linear combinations of the $\hat{c}_\alpha$'s; also, their action
on vacuum on the right gives zero. Similarly, the $\hat{b}_i^+$'s
anticommute among themselves and their action on vacuum on the left gives zero.
Finally, one derives from Eq.(\ref{eq:adj}) the anticommutation relations
\be
\{\hat{b}_i,\hat{b}_j^+\} = \delta_{i,j}.
\ee
We can then construct a Fock basis as
\be
|\{n\}\rangle \equiv \left(b_1^+\right)^{n_1} \ldots
\left(b_M^+\right)^{n_M} |0\rangle
\ee
where $M$ is the order of the matrix $\cal C$, that is the dimension
of the single particle state space, and the occupation numbers $n_i$
are equal to 0 or 1.
The adjoint Fock basis is 
\be
\overline{\langle \{n\}|} \equiv \langle 0| 
\left(b_M\right)^{n_M} \ldots \left(b_1\right)^{n_1}.
\ee

The advantage of these algebraic constructions is that the operator
$C$ has now simple expressions and actions:
\be
C =\sum_i {\cal C}_i \hat{b}_i^+ \hat{b}_i
\ee
and, from the usual calculus in second quantized form:
\be
C |\{n\}\rangle = \left(\sum_i {\cal C}_i n_i\right) |\{n\}\rangle.
\ee
As a consequence
\be
\mbox{Tr}\,\left[e^C\right] = \sum_{\{n\}} \prod_i e^{n_i {\cal C}_i}
=\mathrm{det}\,\left(1+e^{\cal C}\right).
\ee
One has also
\be
\frac{\mbox{Tr}\,\left[\hat{b}_i^+\hat{b}_j e^C\right]}
{\mbox{Tr}\,\left[e^C\right]}= \delta_{i,j} \frac{1}{1+e^{-{\cal C}_i}}
\ee
and finally, using the closure relation Eq.(\ref{eq:clos}), we get
Eq.(\ref{eq:onebodyc}).

\subsection{Back to $g_1$}
We assume now that the matrix $\cal A$ takes the specific form
$i\pi {\cal Q}$ where the matrix 
${\cal Q}$ is a projector on the sites in between $x$ and $y$:
\be
{\cal Q}_{\alpha\beta} =\delta_{\alpha,\beta} \sum_{z=x}^{y}
\delta_{\alpha,z}.
\label{eq:projq}
\ee
This leads to strong simplifications.
First one has
\be
e^{\cal A} = 1 - 2 {\cal Q}.
\ee
So, from Eq.(\ref{eq:matabc}):
\be
1+e^{\cal C} =1+ (1-2 {\cal Q})e^{\cal B} = \left[1-2 {\cal Q} {\cal S}\right]
(1+e^{\cal B}) 
\ee
where we have introduced
\be
{\cal S} \equiv \frac{1}{1+e^{-{\cal B}}}.
\ee
This last matrix has a very simple physical interpretation: using
Eq.(\ref{eq:onebodyc}) with $C$ replaced by $B$, we find that
${\cal S}_{\alpha\beta}$ is the thermal average of $\hat{c}_\beta^\dagger
\hat{c}_\alpha$.
We finally obtain
\be
\frac{1}{1+e^{-{\cal C}}} = e^{\cal C} \frac{1}{1+e^{\cal C}}=
(1 - 2 {\cal Q}) {\cal S}\frac{1}{1-2 {\cal Q} {\cal S}}.
\ee
The matrix to invert in the above expression can be put in block
triangular form by collecting the lattice sites in between $x$ and $y$
in one block, and the other lattice sites in a second block.
In block notations, one can check that
\be
\left(
\begin{array}{cc}
\mathrm{block}_{11} & \mathrm{block}_{12} \\
0 & \mathrm{block}_{22}
\end{array}
\right)^{-1} = 
\left(
\begin{array}{cc}
\mathrm{block}_{11}^{-1} & -\mathrm{block}_{11}^{-1}\,\mathrm{block}_{12}\,
\mathrm{block}_{22} \\
0 & \mathrm{block}_{22}^{-1}                     
\end{array}
\right).
\ee
Here the right upper block of the inverse matrix is not useful as the lattice
site $y$, on which the inverse matrix acts in the expression for
$g_1(x,y)$, belongs to the first block. One has also that the determinant
of a block triangular matrix is the product of the determinants of
block$_{11}$ and block$_{22}$. This leads to
\be
g_1(x,y) =  \frac{1}{2l}\left[-\delta_{x,y}+
\left(\frac{\cal Q}{{\cal Q}-2{\cal Q}{\cal S}{\cal Q}}\right)_{x,y}\right]
\mathrm{det}\, \left[{\cal Q}-2{\cal Q}{\cal S}{\cal Q}\right].
\ee
When specialized to the case $x\neq y$, this gives Eq.(\ref{eq:g1kor}).


\begin{thebibliography}{99}
\bibitem{normali}
Note that the normalization factor of $P$ is infinite for a system with an infinite
number of modes. This formal problem is solved by the introduction
of an energy cut-off that one puts to infinity at the end of the calculations.
\bibitem{intfunc}
We recall that a functional integral is defined as a limit 
of a discrete representation. Here, one should first discretize
the `time' $z$ with a step $dz$, and then take the limit $dz\rightarrow 0$.
This implies that the Laplacian in the energy functional $E[\psi]$ 
should be considered as the limit of a discrete Laplacian, e.g. with
the usual 3-point formula.
The cyclic boundary condition for the discrete Laplacian ensures
that its eigenmodes at a given energy are two counterpropagating
plane waves. This would not be the case for the continuous
Laplacian, in the absence of the extra condition 
$\partial_z \psi(0)=\partial_z \psi(L)$.
\bibitem{winding_number}
More rigorously, one has also to perform a sum over all possible winding numbers
of the field. In the thermodynamic limit this has no effect on observables
like the first order or second order correlation functions considered here,
but it has a dramatic effect on the superfluid fraction, see \cite{super1,super2}
\bibitem{arbitrary}
In particular, there is no relation between $E$, $\hbar^2 k^2/m$
and $\hbar^2 {k'}^{2}/m$.
\bibitem{premier_terme}
The first term in the right hand side of this expression 
should not be forgotten, e.g. in symmetry breaking Bogoliubov approaches,
as emphasized in \cite{Blaizot}, Eq.(3.48). 
We have used the identity
$N_0d\mu_0/dN_0 = l^d\sum_{\rb} g_0\rho_0 \partial_{N_0}\rho_0$
to obtain it, see Eq.(105) of \cite{Mora}.
\bibitem{erratum}
The estimate given in \cite{Mora}
that this distance is the largest of the two distances $\lambda$
and $\xi$ is actually wrong.
\bibitem{physics}
In practice it is difficult to exactly satisfy this condition.
One may consider the 1D coupling constant to be infinite 
if, for the maximum wavevector $k_{\rm max}$ relevant in the physical state,
the scattering amplitude is very close to the one of the infinite delta.
This implies $k_{\rm max} a_1 \ll 1$, see Eq.(\ref{eq:T1d}).
\bibitem{difficile} 
In the case where the trapping potential is replaced by
periodic boundary conditions on the bosons with spatial period $L$, 
the point of the grid of position $L$ is identified with the point of position
$0$, so that $\hat{a}_0 = \hat{a}_L$. 
In the fermionic counterpart one finds that
$\hat{c}_L= e^{i\pi \hat{N}} \hat{c}_0$ where $\hat{N}$ gives the total
number of particles. The boundary conditions for the fermions are therefore
periodic for $N$ odd or antiperiodic for $N$ even.
We shall ignore here such point, as the numerical examples to come
correspond to the thermodynamic limit. This point may become important
for finite size systems, and it is explained in an appendix of \cite{stochGutz}
how to deal with it analytically.
\bibitem{concavity}
As the coefficient of $T^3$ is positive, the variance is a convex function of
temperature at $T\ll T_F$. As the variance tends to one at large temperature,
it necessarily presents an inflexion point, that we find numerically to
be at $T/T_F \sim 0.2$.
\bibitem{help}
If $x>y$, one may use the relation $g_1(x,y)=g_1^*(y,x)$.
\bibitem{Lesovik}
It is shown with the bozonisation technique that the distribution is
indeed Gaussian in the large $k_F X$ limit: see the paragraph below
Eq.(31) in the paper by
L. S. Levitov, H.-W. Lee, and G. B. Lesovik, J. Math. Phys.
{\bf 37}, 4845 (1996).
\end{thebibliography}
\end{document}